\documentclass[a4paper,11pt]{article}
\pdfoutput=1 % if your are submitting a pdflatex (i.e. if you have
\usepackage{jheppub} % for details on the use of the package, please
\usepackage{subfiles}
\usepackage{multirow}
\usepackage{comment}

\usepackage[utf8]{inputenc}

\newcommand{\iso}{\cong}

\newcommand{\iu}{{i\mkern1mu}}
\DeclareMathOperator{\tr}{tr}
\DeclareMathOperator{\Tr}{Tr}

\DeclareMathOperator{\diag}{diag}
\DeclareMathOperator{\Pf}{Pf}

\DeclareMathOperator{\Out}{Out}
\DeclareMathOperator{\rank}{rank}
\DeclareMathOperator{\Sym}{Sym}

\DeclareMathOperator{\PE}{\mathrm{PE}}
\DeclareMathOperator{\PLog}{\mathrm{PLog}}

\DeclareMathOperator{\exponents}{exponents}

\newcommand{\HS}{\textrm{HS}}
\newcommand{\qbf}{\mathfrak{q}}
\newcommand{\tHS}{\mathfrak{t}} %HS parameter
\newcommand{\kgg}{n}
\newcommand{\newdot}{+}

\title{\boldmath 4d $\mathcal{N}=3$ indices via discrete gauging}

\preprint{DESY 18-050}

\author{Thomas Bourton,}
\author{Alessandro Pini}
\author{and Elli Pomoni}

\affiliation{DESY Theory Group, Notkestra{\ss}e 85, 22607 Hamburg, Germany}

\emailAdd{thomas.bourton@desy.de}
\emailAdd{alessandro.pini@desy.de}
\emailAdd{elli.pomoni@desy.de}

\abstract{

\bigskip

A class of 4d $\mathcal{N}=3$ SCFTs can be obtained from gauging a discrete subgroup of the global symmetry group of $\mathcal{N}=4$ Super Yang-Mills theory. 
This discrete subgroup contains elements of both the $SU(4)$ R-symmetry group and the $SL(2,\mathbb{Z})$ S-duality group of $\mathcal{N}=4$ SYM. 
We give a prescription for how to perform the discrete gauging at the level of the superconformal index and Higgs branch Hilbert series.
We interpret and match the information encoded in these indices to
known results for rank one $\mathcal{N}=3$ theories. Our prescription is easily generalised for the Coulomb branch and the Higgs branch indices of higher rank theories, allowing us to make new predictions for these theories. Most strikingly we find that the Coulomb branches of higher rank theories are generically not-freely generated.

\bigskip

}

\begin{document} 
\maketitle
\flushbottom

\newpage

\section{Introduction}

%The study of supersymmetric theories has taught us a lot about Quantum Field theory and especially non-perturbative phenomena as the extra symmetry they possess
%allows for exact computations. 
In recent years a lot of insight has been gained by trying to understand the landscape of supersymmetric quantum field theories and especially superconformal field theories (SCFTs). Even though much progress has been made towards understanding the properties of $\mathcal{N} = 1,2,4$ theories, in four dimensions,
the $\mathcal{N} = 3$ case has been long ignored.
This is due to the fact that up until very recently no example of a genuinely $\mathcal{N} = 3$ theory was known.
%and most of the computational tools available require a  Lagrangian description in some frame where the theory is weakly coupled.
Moreover, the only multiplet of $\mathcal{N} = 3$ supersymmetry that can be free is a vector
multiplet which, after imposing CPT invariance, is identical to the $\mathcal{N}=4$ vector multiplet. Thus, there are no free genuinely $\mathcal{N} = 3$ theories and all genuinely $\mathcal{N} = 3$ theories have to be strongly coupled.

%In recent years a lot of mileage has been gained by trying to understand the landscape of supersymmetric quantum field theories and especially superconformal field theories (SCFTs).
The first to seriously consider  the  consequences  of $\mathcal{N} = 3$ supersymmetry
was
\cite{Aharony:2015oyb}
who, via the study of
$\mathcal{N} = 3$ superconformal symmetry, were able to reveal several basic properties, which consistent $\mathcal{N} = 3$ theories should possess, if they exist.
These properties include the
fact that these SCFTs have no marginal couplings and are therefore isolated fixed points. This is to be contrasted with generic $\mathcal{N}=2$ and $\mathcal{N}=4$ gauge theories, which have a conformal manifold parametrised by the complexified gauge couplings.
%(usually $\mathcal{N} = 2$ SCFTs have a conformal manifold parametrised by marginal couplings)
Additionally, the conformal anomalies $a$ and $c$ must be equal, as is the case for $\mathcal{N} = 4$ theories, while, for generic $\mathcal{N}=2$ theories $a\neq c$.

Moreover,  $\mathcal{N} = 3$ SCFTs cannot have a flavour symmetry that is not an R-symmetry, as is the case for $\mathcal{N}=4$ theories.
Finally,  some basic properties of  the infrared physics can be extracted from the study of the supersymmetric vacua.
Seen as $\mathcal{N} = 2$ theories, $\mathcal{N} = 3$ theories have a Coulomb and a Higgs branch, with the two branches  related to each other by the $SU(3)$ R-symmetry of the $\mathcal{N} = 3$ superconformal algebra.

Garc{\'i}a-Etxebarria and Regalado \cite{Garcia-Etxebarria:2015wns} were the first to discover/construct examples of $\mathcal{N} = 3$ theories embedded in type IIB string theory (F-theory) by generalizing the well known orientifold construction to $\mathcal{N} = 3$ preserving S-folds.
The S-fold includes a $\mathbb{Z}_k$ projection on both the R-symmetry directions as well as the $SL(2,\mathbb{Z})$
S-duality group of type IIB (the torus of F-theory).
The list of known $\mathcal{N} = 3$ theories was then further enhanced by \cite{Aharony:2016kai}
via a classification of different variants of S-folds, distinguished by an analog of discrete torsion.
Moreover, \cite{Aharony:2016kai} also clarified the role of discrete gauge and global symmetries.
Finally, a path to the construction of even more $\mathcal{N}=3$ SCFTs was given in \cite{Argyres:2016yzz} via gauging a discrete subgroup of the global symmetry group of $\mathcal{N}=4$ SYM.

$\mathcal{N}=3$ SCFTs, exactly because they do not have a Lagrangian description, can be studied only with certain tools. Representation theory alone can take us very far \cite{Aharony:2015oyb,Cordova:2016xhm,Evtikhiev:2017heo}.
String theory, F-theory and M-theory constructions provide the primary way that we have to study $\mathcal{N}=3$ SCFTs \cite{Garcia-Etxebarria:2015wns,Garcia-Etxebarria:2016erx}. The type IIB description allows for an AdS gravity dual description which can be used to examine the properties of $\mathcal{N}=3$ theories in the large $N$ limit \cite{Garcia-Etxebarria:2015wns,Aharony:2016kai,Imamura:2016abe}. Moreover, $\mathcal{N}=3$ theories have Seiberg-Witten solutions \cite{Seiberg:1994aj,Seiberg:1994rs} which encode the low energy effective action of the theory on the Coulomb branch. Various aspects of the Coulomb branches for these theories have been studied in \cite{Argyres:2015ffa,Argyres:2015gha,Argyres:2016xua,Argyres:2016xmc,Aharony:2016kai,Aharony:2015oyb,Argyres:2016yzz,Argyres:2018wxu}.
Another powerful tool is the superconformal bootstrap which has been studied in \cite{Lemos:2016xke}.
The bootstrap can also be suplemented with chiral algebra techniques \cite{Beem:2013sza} and has been studied in \cite{Nishinaka:2016hbw,Lemos:2016xke}. Further techniques have been developed in \cite{Imamura:2016udl,Agarwal:2016rvx}.

In this paper we take the path of the superconformal index. Usually the superconformal index can be computed only for theories with a Lagrangian description, where one may take a free field limit and use letter counting. Genuine $\mathcal{N}=3$ SCFTs do not admit a free field limit and therefore it is not possible to use standard techniques.
In \cite{Imamura:2016abe} the superconformal index was computed in the large $N$ limit via matching it with the KK reduction of the gravity dual of the $\mathcal{N}=3$ SCFT.
Here we follow another path, that will lead to the answer for any $N$, inspired by the ``orbifolding procedure'' which gives the index of a daughter theory from a mother that we recently used in \cite{Bourton:2017pee}.
Based on the observation that certain $\mathcal{N}=4$ SYM theories have an enhanced discrete global symmetry at certain values of the gauge coupling, we point out that the superconformal index may be refined by a further fugacity for the enhanced discrete symmetry.   
The index of the discretely gauged daughter theory is then obtained by ``integrating'' over the additional fugacity $\epsilon$, which takes values in the discrete group. Schematically,
\begin{equation}
\mathcal{I}_{\mathcal{N}=3}=\frac{1}{|\mathbb{Z}_{\kgg}|}\sum_{\epsilon\in\mathbb{Z}_{\kgg}}\mathcal{I}_{\mathcal{N}=4}(\epsilon)\,.
\end{equation}

\vspace{0.1cm}

This paper is organized as follows. In Section \ref{sec:N3Cons} we review the possible constructions of $\mathcal{N}=3$ SCFTs via S-folding and via discrete gauging. This gives us the opportunity to discuss in detail the symmetries of both the mother and the daughter theories and to embed the discrete subgroup that we want to gauge in the $SU(4)$ R-symmetry group and the $SL(2,\mathbb{Z})$ S-duality group of $\mathcal{N}=4$ SYM. 
In Section \ref{sec:su22Nrepn} we gather some facts about representation theory of $\mathfrak{su}(2,2|\mathcal{N})$ superconformal algebras that we will need for the index computation and interpretation.
In Section \ref{sec:Index} we introduce the refined version of the superconformal index, its Coulomb branch limit and the Higgs branch Hilbert series. The discrete gauging prescription is presented and the procedure for computing it is introduced.
Section \ref{sec:rank1} is devoted to rank one examples. 
Section \ref{sec:rankN} deals with higher rank examples. We focus on the Coulomb and Higgs branches. Our higher rank computations allow us to make new predictions for these theories. 
Finally, in Section \ref{sec:largeN} we compute the single trace index in the large $N$ limit and match to the AdS/CFT result of \cite{Imamura:2016abe}.
 
\bigskip
\textbf{Note added:} While this paper was being completed we became aware of \cite{Argyres:2018wxu}, with which, although our methods are different, there is considerable overlap with our results. In particular, that paper also describes $\mathcal{N}=3$ theories obtained via discrete gauging of $\mathcal{N}=4$ SYM. In the cases where our results overlap, they agree. We would like to thank P. Argyres and M. Martone for sharing the draft and for discussing their results with us. There is also some overlap with \cite{Antoine} which appeared while we were finishing writing the paper.

\section{\boldmath  Constructing the $\mathcal{N}=3$ theories}
\label{sec:N3Cons}
\subsection{S-folds}\label{sec:sfold}
\label{N3Cons}
One possible way to realise $\mathcal{N}=3$ SCFTs is via S-folds. S-folds were originally introduced in \cite{Garcia-Etxebarria:2015wns} and are non-perturbative generalisations of the standard orientifolds in string theory. The construction introduced in \cite{Garcia-Etxebarria:2015wns} goes as follows: consider F-theory on $\mathbb{R}^4\times\left(\mathbb{R}^6\times T^2\right)/\mathbb{Z}_k$. The $\mathbb{Z}_k\subset Spin(6)\times SL(2,\mathbb{Z})$ and we denote its generator in $\mathfrak{su}(4)\iso\mathfrak{so}(6)$ to be $r_k$ which acts on the coordinates $X^i$, $i=1,\dots,6$ of $\mathbb{R}^6$ by rotation corresponding to
\begin{equation}
R_k=e^{\frac{2\pi\iu}{k} (q_1+q_2-q_3)}=\begin{pmatrix}
\hat{R}_k&0&0\\
0&\hat{R}_k&0\\
0&0&\hat{R}_k^{-1}
\end{pmatrix}\in SO(6)\,,
\end{equation}
where $\hat{R}_k$ denotes rotation by $2\pi/k$ in the corresponding 2-plane. $q_1$, $q_2$, $q_3\in\mathfrak{so}(6)$ denote the Cartan generators of $\mathfrak{so}(6)$. The corresponding element in $SU(4)\iso Spin(6)$ is just
\begin{equation}\label{eqn:Rsym}
\widetilde{R}_k=e^{\frac{2\pi\iu}{k}r_k}=e^{\frac{2\pi\iu}{k} (\frac{R_1}{2}+R_2+\frac{3R_3}{2})}=\begin{pmatrix}
e^{\iu\pi/k}&0&0&0\\
0&e^{\iu\pi/k}&0&0\\
0&0&e^{\iu\pi/k}&0\\
0&0&0&e^{-3\iu\pi/k}
\end{pmatrix}\in SU(4)\,.
\end{equation}
We choose a basis for the Cartans $R_1$, $R_2$, $R_3\in\mathfrak{su}(4)$ given by\footnote{We use the same conventions as in \cite{Kinney:2005ej} and the $\mathfrak{so}(6)$ Dynkin labels $(q_1,q_2,q_3)$ are related to the $\mathfrak{su}(4)$ Dynkin labels $(R_1,R_2,R_3)$ by
\begin{equation*}
q_1=\frac{R_1}{2}+R_2+\frac{R_3}{2}, \ \ q_2=\frac{R_1}{2}+\frac{R_3}{2}, \ \ q_3=\frac{R_1}{2}-\frac{R_3}{2}\,.
\end{equation*}
} 
\begin{equation}
R_1=\diag(1,-1,0,0)\,,\quad R_2=\diag(0,1,-1,0)\,,\quad R_3=\diag(0,0,1,-1)\,.
\end{equation}

On the other hand the quotient on the torus acts as an involution of the torus only for $k=1,2,3,4,6$. Moreover $k=3,4,6$ require fixed complex structure of $\tau=e^{\iu\pi/3},\iu,e^{\iu\pi/3}$ respectively. In that case we denote the generator of $\mathbb{Z}_k\subset SL(2,\mathbb{Z})$ by $s_k$. $s_k$ acts on the coordinate $x+\tau y$ of the $T^2$ corresponding to $S_k\in SL(2,\mathbb{Z})$ with
\begin{equation}\label{eqn:Sduality}
S_2=\begin{pmatrix}
-1&\phantom{-}0\\
\phantom{-}0&-1
\end{pmatrix}\,,\quad S_3=\begin{pmatrix}
\phantom{-}0&\phantom{-}1\\
-1&-1
\end{pmatrix}\,,\quad S_4=\begin{pmatrix}
\phantom{-}0&\phantom{-}1\\
-1&\phantom{-}0
\end{pmatrix}\,,\quad S_6=\begin{pmatrix}
0&-1\\
1&\phantom{-}1
\end{pmatrix}\,.
\end{equation}
The elements of $\mathbb{Z}_k\subset Spin(6)\times SL(2,\mathbb{Z})$ are of the form $e^{\frac{2\pi\iu}{k}(r_k+ s_k)}$ , corresponding to the combined action \eqref{eqn:Rsym} and \eqref{eqn:Sduality}.

After taking the type IIB limit of F-theory the singular geometry can be probed with a stack of $N$ D3-branes. The resulting low energy theory on the D3-branes (for $k=3,4,6$) is a strongly interacting $\mathcal{N}=3$ SCFT. In Appendix \ref{sec:preservedSCA} we explicitly show the supercharges that are preserved by the $\mathbb{Z}_k$ quotient.

A careful analysis \cite{Aharony:2016kai} of the discrete global symmetries indicates, as for the $(k=2)$ $O3^{\pm},\widetilde{O3}^{\pm}$ perturbative orientifolds, the $k=3,4,6$ S-folds are characterised by different $\mathbb{Z}_p\subset\mathbb{Z}_k$ global symmetries. The S-fold variants are then labelled by $k,\ell=k/p$. We denote the theory of $N$ D3-branes by $S^N_{k,\ell}$ and it has Coulomb branch operators of dimension
\begin{equation}
k\,,\,2k\,,\,\dots\,,\,(N-1)k\,;\,N\ell \, ,
\end{equation}
corresponding to Coulomb branch operators $\left(\sum_{i=1}^Nz^{jk}_i\right)$, $j=1,\dots,N-1$, and the Pfaffian-like operator $(z_1z_2\dots z_N)^{\ell}$ where $z_i$ denote the positions of the D3-branes in $\mathbb{C}/\mathbb{Z}_k$. Consequently the theory has central charge given by \cite{Argyres:2007tq,Shapere:2008zf,Aharony:2016kai}
\begin{equation}\label{eqn:Sfoldac}
a_{k,\ell}=c_{k,\ell}=\frac{kN^2+(2\ell-k-1)N}{4}\,.
\end{equation}
The theory $S_{k,\ell}^N$ associated to each value of $k$, $\ell$ has a global symmetry of (at least) $\mathbb{Z}_p=\mathbb{Z}_{k/\ell}$ which acts on the Pfaffian-like operator $(z_1z_2\dots z_N)^{\ell}\mapsto (e^{2\pi\iu/k} z_1z_2\dots z_N)^{\ell}=e^{2\pi\iu/p}(z_1z_2\dots z_N)^{\ell}$ while acting trivially on every other Coulomb branch operator. By gauging $\mathbb{Z}_{p'}\subset\mathbb{Z}_{p}\subset\mathbb{Z}_k$ discrete symmetry we obtain further theories 
\begin{equation}\label{eqn:sfolddisgauge}
S^N_{k,\ell}\xrightarrow{\text{$\mathbb{Z}_{p'}$ gauging}} S^N_{k,\ell,p'} \, ,
\end{equation} 
which, since they arise as discrete gauging of a `parent' theory, have central charge \eqref{eqn:Sfoldac} and the theory $S^N_{k,\ell,p'}$ has Coulomb branch operators of dimension
\begin{equation}\label{eqn:sfolddiscgauge}
k\,,\,2k\,,\,\dots\,,\,(N-1)k\,;\,Np'\,.
\end{equation}
Since the $\mathbb{Z}_{p'}$ acts non-trivially only on a single operator quotienting by $\mathbb{Z}_{p'}$ does not introduce relations and the corresponding ring is freely generated.

\subsection{$\mathcal{N}=3$ preserving discrete gauging}
In this paper we use a different construction to the one described in Section \ref{sec:sfold}. Consider instead $\mathcal{N}=4$ SYM with gauge group $G$. The theory has an exactly marginal gauge coupling $\tau$. $\mathcal{N}=4$ SYM (on $\mathbb{R}^4$) has an S-duality group generated by \cite{MONTONEN1977117,GNO,Vafa:1994tf,Argyres:2006qr,Dorey:1996hx,Girardello:1995gf,Kapustin:2006pk}
\begin{equation}\label{eqn:Sdual}
\left(\tau\,,G\right)\mapsto \left(\tau+1\,,G\right)\,\quad \text{and} \quad\left(\tau\,,G\right)\mapsto\left(-\frac{1}{\lambda_q^2\tau}\,,{}^LG\right)\, ,
\end{equation}
where $\lambda_q=2\cos\frac{\pi}{q}$ and ${}^LG$ denotes the Langlands dual of $G$. The action on $\tau$ forms a group known as the Hecke group $H(\lambda_q)\subset SL(2,\mathbb{Z}\left[\lambda_q\right])$ and it is generated by
\begin{equation}
T=\begin{pmatrix}
1&1\\
0&1
\end{pmatrix}\,,\quad S =\begin{pmatrix}
0&-\lambda_q^{-1}\\
\lambda_q&0
\end{pmatrix}\,.
\end{equation}
For $q=3$ $H(\lambda_3)=H(1)=SL(2,\mathbb{Z})$. Let $\mathfrak{g}=Lie(G)$. When $\mathfrak{g}=ADE$ (or $\mathfrak{u}(N)$) $q=3$ while for $\mathfrak{g}=BCF$ $q=4$ and for $\mathfrak{g}=G_2$ $q=6$.
\begin{comment} It can be shown that
\begin{equation}\label{eqn:Heckegroup}
H(\lambda_q)\iso\langle s,t,w|s^2=t^q=w,w^2=1\rangle \, ,
\end{equation}
with the isomorphism made by $s=S$, $t=ST$ \cite{singerman1998,ikikardes2006,LANG2000220}.\footnote{Generally Hecke groups are usually discussed for $\tau\in\mathbb{H}$ in that case they are related to our discussion as $H'(\lambda_q)=H(\lambda_q)/\langle w\rangle$, then $H'(\lambda_3)\iso PSL(2,\mathbb{Z})$. When we say Hecke group we always mean the $\mathbb{Z}_2$ central extension $0\to H'(\lambda_q)\to H(\lambda_q)\to\mathbb{Z}_2\to0$.} 
\end{comment}
We define the \textit{self-duality group} of the theory with gauge group $G$ to be the subset of transformations $\tau\mapsto\tau'$ in $H(\lambda_q)$ which map the theory to itself. When one considers non-local operators this subset of transformations is generally a subgroup of $H(\lambda_q)$ due to the fact that $G\mapsto{}^LG$ clearly changes the global structure of the theory and therefore the spectrum of non-local operators. However, at the level of local operators,    
when $\mathfrak{g}=ADE$ (or $\mathfrak{u}(N)$) we have $\mathfrak{g}={}^L\mathfrak{g}$ and then at the level of local operators the \textit{self-duality group} is simply the full $SL(2,\mathbb{Z})$. On the other hand when $\mathfrak{g}=BCFG$ then $\mathfrak{g}\neq{}^L\mathfrak{g}$. In particular $B_N\neq{}^LB_N\iso C_N$, $F_4\neq{}^LF_4\iso F_4$ and $G_2\neq{}^LG_2\iso G_2$\footnote{S-duality transformations for non-simply-laced Lie algebras are more complicated. In the particular cases of $G_2$ and $F_4$ Lie algebras we can perform a rotation on the root system of the corresponding Langland dual algebras $^{L}G_2 = G_2^{'}$ and $^{L}F_4 = F_{4}^{'}$ such that these turn out to be isomorphic to the initial one \cite{Argyres:2006qr}.}

 and the \textit{self-duality group} even at the level of local operators is reduced to a subgroup of $H(\lambda_q)$.\footnote{The Langlands dual algebra is obtained by exchanging $\alpha\mapsto\alpha^{\vee}=\frac{2}{(\alpha\cdot\alpha)}\alpha$. For simply laced algebras we have $\alpha^{\vee}=\alpha$ and $\mathfrak{g}={}^L\mathfrak{g}$. On the other hand, when $\mathfrak{g}$ is not simply laced $\alpha^{\vee}\neq\alpha$ if $\alpha$ is a long root and $\mathfrak{g}\neq{}^L\mathfrak{g}$.} In this paper we will discuss only the cases when $\mathfrak{g}={}^L\mathfrak{g}$.
% We should also point out that the insertion of non-local operators generally reduces the full $H(\lambda_q)$ to a subgroup due to the exchange $G\to{}^LG$.
Let us now discuss the possible symmetry enhancements. $SL(2,\mathbb{Z})$ has finite cyclic subgroups
\begin{equation}\label{eqn:taufix}
\text{$SL(2,\mathbb{Z})\supset \mathbb{Z}_{\kgg}$ for $\kgg=\{2,3,4,6\}$ that fixes $\tau=\left\{\text{any},e^{\iu\pi/3},\iu,e^{\iu\pi/3}\right\}$}\,,
\end{equation}
where we take only those fixed points with $\text{Im }\tau\geq0$. The $\mathbb{Z}_{\kgg}$ are generated by the $S_{\kgg}$ as in equation \eqref{eqn:Sduality}
\begin{equation}
S_2=S^2\,,\quad S_3=S^3T\,,\quad S_4=S^3\,,\quad S_6=ST\,.
\end{equation}
\begin{comment}
$H(\lambda_q)$ has finite cyclic subgroups given by
\begin{equation}\label{eqn:taufix}
\text{$H(\lambda_q)\supset \mathbb{Z}_{\kgg}$ for $\kgg=\{2,4,q,2q\}$ fixes $\tau=\left\{\text{any},\frac{\iu}{2\cos\frac{\pi}{q}},-\frac{1}{2}+\frac{\iu}{2}\tan\frac{\pi}{q},-\frac{1}{2}+\frac{\iu}{2}\tan\frac{\pi}{q}\right\}$}\,,
\end{equation}
where we take only those fixed points with $\text{Im }\tau\geq0$. For $q=3$ these of course reduce to the usual order $\kgg=4,3,6$ $SL(2,\mathbb{Z})$ fixed points at $\tau=\iu,e^{\iu\pi/3},e^{\iu\pi/3}$.
\end{comment}

At a generic point on the conformal manifold the global symmetry group of the theory is at least $PSU(2,2|4)$.
%\footnote{There may also be the $\mathbb{Z}_2\subset SU(4)\times H(\lambda_q)$ generated by $r_2\cdot s_2$ which preserves $\mathcal{N}=4$ supersymmetry and does not fix a value of $\tau$. The $\mathbb{Z}_2$ acts on the entire $\mathcal{N}=4$ vector multiplet as $s_2\cdot r_2:V_{\mathcal{N}=4}\mapsto-V_{\mathcal{N}=4}$. However in some cases, depending on the choice of $G$, this $\mathbb{Z}_2$ is actually isomorphic to conjugation by an element $g\in G$, i.e. $g^{-1}V_{\mathcal{N}=4}g=-V_{\mathcal{N}=4}$ in which case the $\mathbb{Z}_2$ does not generate a global symmetry since it is isomorphic to a gauge transformation.}
On the other hand, for $\tau$ fixed as in \eqref{eqn:taufix}, the global symmetry group (acting on local operators) has a $\mathbb{Z}_n$  enhancement
%correspondingly enhances to (at least) $PSU(2,2|4)\times\mathbb{Z}_{\kgg}$ 
for ${\kgg}=3,4,6$ where the $\mathbb{Z}_{\kgg}$ is generated by \eqref{eqn:Sduality}. We use the notation $\mathbb{Z}_{\kgg}$ since
%, as we will discuss below,
${\kgg}$ should generally be considered unrelated to the parameters $k,\ell,p'$ appearing in the S-fold construction of the previous section. We therefore have a discrete global symmetry
\begin{equation}\label{eqn:Zkgauge}
\mathbb{Z}_{\kgg}\subset SU(4)\times SL(2,\mathbb{Z})
\end{equation}
generated by $r_{\kgg}+ s_{\kgg}$. We may consider gauging the $\mathbb{Z}_{\kgg}$ (or in the case when ${\kgg}$ is not prime, subgroups of the $\mathbb{Z}_{\kgg}$) global symmetry \cite{Argyres:2016yzz}. Doing so results in a new theory with a different spectrum of local and non-local operators, but, with equivalent local dynamics and therefore the same values for the $a$ and $c$ anomaly coefficients. 
%One may also worry that this $\mathbb{Z}_{\kgg}$ may have non-vanishing 't Hooft anomalies. We give an argument that this is not the case (at least for the topologically twisted theory on $\mathbb{S}^1\times \mathbb{S}^3$) in Appendix \ref{Sec:thooftanomaly}. 
The action \eqref{eqn:Zkgauge} preserves the same supercharges as the $\mathbb{Z}_k$ S-fold, i.e. the ${\kgg}=3,4,6$ discrete gaugings preserves four dimensional $\mathcal{N}=3$ supersymmetry.
Therefore, the theories we will construct are to be labelled by the parent $\mathcal{N}=4$ theory and the discrete group to be gauged. The possible parent theories are labelled by a choice of gauge group $G$. We will only consider parent theories where $G$ is connected. 

Moreover, since we will eventually be interested in computing quantities sensitive only to the local operator spectrum, the global form of the gauge group will not play a role in the computations\footnote{Since $\pi_1\left(\mathbb{S}^3\right)=\pi_2\left(\mathbb{S}^3\right)=\{1\}$ the superconformal index ($\mathbb{S}^3\times\mathbb{S}^1$ partition function) is sensitive only to the spectrum of local operators i.e., for connected groups, a choice of Lie algebra $\mathfrak{g}$.} and therefore the theories should be rather be labelled by the choice of Lie algebra $\mathfrak{g}$ of $G$. 

\paragraph{Coulomb branch}
Let us now briefly compare with the construction in the previous subsection. Considered as an $\mathcal{N}=2$ theory we have algebraically independent (over $\mathbb{C}$) Coulomb branch operators $u_j$, $1\leq j\leq N$ , $N:=\rank\mathfrak{g},$ of dimension $E(u_j)$. In the notation of \cite{DolanOsborn} the $u_j$'s are the highest weight states of the chiral $\mathcal{E}_{r,(0,0)}$ multiplets, with conformal dimension $E(u_j)=r(u_j)$ where $r(u_j)$ is the charge under the $\mathfrak{u}(1)_r$ of the $\mathcal{N}=2$ superconformal algebra (see Table \ref{tab:short}). They are built up out of $\mathfrak{g}$-invariant combinations of the scalar $X\in\mathfrak{h}$ in the $\mathcal{N}=2$ vector multiplet, where $\mathfrak{h}$ is a Cartan subalgebra of $\mathfrak{g}$, while setting the adjoint hypermultiplet scalars $Y=Z=0$. Let us now go to a point on the conformal manifold where we have an enhanced $\mathbb{Z}_{\kgg}$ global symmetry generated by $r_{\kgg}+ s_{\kgg}$. In comparison with the discussion \eqref{eqn:Sfoldac}-\eqref{eqn:sfolddiscgauge} this $\mathbb{Z}_{\kgg}$ global symmetry acts non-trivially on multiple Coulomb branch operators of the parent theory, namely
\begin{equation}\label{eqn:disgaugeZnaction}
\mathbb{Z}_{\kgg}: u_j\mapsto e^{\frac{2\pi\iu}{\kgg} E(u_j)}u_j\,.
\end{equation}
It is clear that this $\mathbb{Z}_{\kgg}$ action does not generically generate a complex reflection group $G(\rank\mathfrak{g},m,\kgg)$\footnote{See equation (2.10) of \cite{Aharony:2016kai} for a definition.} on $CB_{\mathfrak{g}}:=\mathbb{C}\left[u_1,u_2,\dots,u_{\rank\mathfrak{g}}\right]$ and therefore, by the Chevalley-Shephard-Todd theorem \cite{shephard1954finite,Aharony:2016kai}, the resulting quotient ring generically has relation(s). %Infact the action \eqref{eqn:disgaugeZnaction} generates a group that is not isomorphic to a complex reflection group if the set of generators $\{u_j\}$ of the Coulomb branch contains two or more operators $u$, $u'$ such that $E(u)\neq0\mod n$ and $E(u')\neq0\mod \kgg$.
Hence, when $\rank\mathfrak{g}\geq2$ and $\kgg\geq2$, the quotient of the Coulomb branch of the parent theory $CB_{\mathfrak{g}}$ by \eqref{eqn:disgaugeZnaction}
\begin{equation}\label{eqn:CBquotient}
CB_{\mathfrak{g},\kgg}:=CB_{\mathfrak{g}}/\mathbb{Z}_{\kgg}
\end{equation}
generally has a non-planar topology. We will see that the structure of the ring can be often be deduced by studying the Coulomb branch index. Some properties of non-freely generated Coulomb branch chiral rings were described in \cite{Argyres:2017tmj}.
We would also like to point out that in \cite{Aharony:2016kai} discrete gauging which results in non-freely generated Coulomb branches was explicitly not considered. They considered discrete gauging of the parent theories $S_{k,\ell}^N$ of only $\mathbb{Z}_{p'}\subset\mathbb{Z}_{k/\ell}$ discrete symmetry which acts non-trivially only on a single Coulomb branch operator. However these theories may have larger discrete symmetry groups which may act non-trivially on multiple Coulomb branch operators. Upon gauging such discrete symmetries one can obtain theories with non-freely generated Coulomb branches.
Because the discrete gauging does not change the values of $a$ and $c$ we expect them to be equal to those of the $\mathcal{N}=4$ parent theory. If the Coulomb branch operators of the $\mathcal{N}=4$ parent theory have dimension $E(u_i)$ then the $a$ and $c$ anomaly coefficients are given by \cite{Argyres:2007tq,Shapere:2008zf}
\begin{equation}\label{eqn:discgaugeac}
a=c=\sum_{i=1}^{\rank \mathfrak{g}}\frac{2E(u_i)-1}{4}\,.
\end{equation}

\paragraph{Higgs branch}
Considered as a $\mathcal{N}=2$ theory the Higgs branch is reached by setting $X=0$ and by giving diagonal vevs to the adjoint hypermultiplet scalars $Y,Z\in\mathfrak{h}$. The Higgs branch $HB_{\mathfrak{g}}$ is then parametrised by $\mathfrak{g}$-invariant combinations $W_{i}^{(f)}$ of the $Y,Z$ that transform in the $f$-representation of $U(1)_f$. Where $U(1)_f$ is the flavour symmetry that all $\mathcal{N}=3$ theories have, when seen as $\mathcal{N}=2$ theories, as we will review in Section \ref{sec:su22Nrepn}. In the notation of \cite{DolanOsborn} the $W_{i}^{(f)}$ are the highest weight states of the $\hat{\mathcal{B}}_R$ multiplets and have $E=2R$ and $r=0$, where $R$ is the Cartan of the $\mathfrak{su}(2)_R$ R-symmetry of the $\mathcal{N}=2$ superconformal algebra (see Table \ref{tab:short}). When $\mathfrak{g}$ is non-abelian $HB_{\mathfrak{g}}$ is generically non-freely generated. Since $Y,Z$ have $s_{\kgg}=0$ and $r_{\kgg}=r+f=f$ the $\mathbb{Z}_{\kgg}$ acts by
\begin{equation}\label{eqn:HBaction}
\mathbb{Z}_{\kgg}:W_{i}^{(f)}\mapsto e^{\frac{2\pi\iu}{\kgg}f}W_{i}^{(f)}\,.
\end{equation}
Therefore, after the discrete gauging, the Higgs branch is given by the quotient of the Higgs branch of the parent theory $HB_{\mathfrak{g}}$ by the $\mathbb{Z}_{\kgg}$ action \eqref{eqn:HBaction}
\begin{equation}\label{eqn:HBdisgauge}
HB_{\mathfrak{g},\kgg}:=HB_{\mathfrak{g}}/\mathbb{Z}_{\kgg}\,.
\end{equation}
In Section \ref{sec:Index} we discuss how to compute the Hilbert series of \eqref{eqn:HBdisgauge}.

\section{\boldmath{$ \mathfrak{su}(2,2|\mathcal{N})$} representation theory}\label{sec:su22Nrepn}
In this section we will describe some basic facts about representations of (the complexification of) $\mathfrak{su}(2,2|\mathcal{N})$ and their decompositions into subalgebras. 

\subsection{$\mathfrak{psu}(2,2|4)\to\mathfrak{su}(2,2|3)$ decomposition}
The superconformal symmetry algebra of 4d $\mathcal{N}=4$ SYM  is given by $\mathfrak{su}(2,2|4)$. Unitary representations of $\mathfrak{psu}(2,2|\mathcal{N})$ are necessarily non-compact. Unitary representations are labelled by $(E,j_1,j_2,R_1,R_2,R_3)$ which label representations under the maximal bosonic subalgebra
\begin{equation}
\mathfrak{u}(1)_E\oplus\mathfrak{su}(2)_1\oplus \mathfrak{su}(2)_2\oplus \mathfrak{su}(4) \subset \mathfrak{psu}(2,2|4)\,.
\end{equation}
Here $E$ labels the conformal dimension, $j_1,j_2$ label spin representations and $R_1,R_2,R_3$ are the Dynkin labels of $\mathfrak{su}(4)$. 

As we discussed in Section \ref{N3Cons}, upon the $\mathbb{Z}_{\kgg}$ discrete gauging $\mathfrak{psu}(2,2|4)$ superconformal symmetry is broken down to $\mathfrak{su}(2,2|3)$ (for ${\kgg}=3,4,6$). Representations of this algebra are labelled by $(E,j_1,j_2,R_1,R_2,r_{\mathcal{N}=3})$ of the maximal compact bosonic subalgebra
\begin{equation}
\mathfrak{u}(1)_E\oplus\mathfrak{su}(2)_1\oplus \mathfrak{su}(2)_2\oplus \mathfrak{su}(3)\oplus\mathfrak{u}(1)_{r_{\mathcal{N}=3}} \subset \mathfrak{su}(2,2|3)\,.
\end{equation}
In particular $\mathfrak{su}(4)\to \mathfrak{su}(3)\oplus \mathfrak{u}(1)_{r_{\mathcal{N}=3}}$.
%The exact subalgebra is given by the centraliser of the generator of \eqref{eqn:Rsym} in $\mathfrak{su}(4)_R$.
The surviving supercharges are simply given by $\mathcal{Q}_{\alpha}^{I=1,2,3}$, $\widetilde{\mathcal{Q}}_{\dot\alpha I=1,2,3}$ and their conjugates. The Cartans of $\mathfrak{su}(3)$ are given by $R_1,R_2$ and $\mathfrak{u}(1)_{r_{\mathcal{N}=3}}$ is generated by 
\begin{equation}
r_{\mathcal{N}=3}=\frac{R_1}{3}+\frac{2R_2}{3}+R_3
\end{equation}
under which the $\mathcal{Q}_{\alpha}^{I=1,2,3}$ have $r_{\mathcal{N}=3}=\frac{1}{3}$ and $\widetilde{\mathcal{Q}}_{\dot\alpha I=1,2,3}$ have $r_{\mathcal{N}=3}=-\frac{1}{3}$. 

One of the most important multiplets of $\mathfrak{psu}(2,2|4)$ are the half-BPS multiplets called $\mathcal{B}_{[0,R_2,0]}^{\frac{1}{2},\frac{1}{2}}$ in the language of \cite{DolanOsborn}. These multiplets obey maximal shortening given by $R_2=E$.
The superconformal primaries of these multiplets are given by single trace operators of the form $\tr\phi^{(I_1J_1}\dots\phi^{I_mJ_m)}$ (see Table \ref{fig:N4vecletters} for conventions) with $(E,j_1,j_2,R_1,R_2,R_3)=(R_2,0,0,0,R_2,0)$. Under $\mathfrak{psu}(2,2|4)\to\mathfrak{su}(2,2|3)$ these multiplets decompose as
\begin{equation}\label{B0m0}
\mathcal{B}_{[0,R_2,0]}^{\frac{1}{2},\frac{1}{2}}\color{black}\iso\bigoplus_{i=0}^{R_2}\hat{\mathcal{B}}_{[R_2-i,i]}\color{black}\,.
\end{equation}
Note that this is a simple consequence of the branching of $\mathfrak{su}(4)\to\mathfrak{su}(3)\oplus\mathfrak{u}(1)_{r_{\mathcal{N}=3}}$
\begin{equation}\label{eqn:su3branching}
\mathbf{[0,R_2,0]}\to\bigoplus_{i=0}^{R_2}\mathbf{[R_2-i,i]}_{\frac{4i}{3}-\frac{2R_2}{3}}\,,
\end{equation}
where the subscript denotes the $\mathfrak{u}(1)_{r_{\mathcal{N}=3}}$ charge.
The multiplets $\hat{\mathcal{B}}_{[R_1,R_2]}\color{black}$ obey the shortening condition $E=R_1+R_2$, $r_{\mathcal{N}=3}=\frac{2}{3}(R_2-R_1)$. The superconformal primary of these multiplets is given by an operator with 
$(E,j_1,j_2,R_1,R_2,r_{\mathcal{N}=3})=\left(R_1+R_2,0,0,R_1,R_2,\frac{2R_2-2R_1}{3}\right)$ corresponding to the decomposition of $\tr\phi^{(I_1J_1}\dots\phi^{I_mJ_m)}$ under the branching \eqref{eqn:su3branching}. 

\subsection{$\mathfrak{su}(2,2|3)\to\mathfrak{su}(2,2|2)$ decomposition}
For practical applications, rather than dealing with $\mathfrak{su}(2,2|3)$ representations, it is often convenient to choose a $\mathfrak{su}(2,2|2)\subset \mathfrak{su}(2,2|3)$ subalgebra. Representations of this algebra are labelled by $(E,j_1,j_2,R,r)$ under the maximal bosonic subalgebra
\begin{equation}\label{eqn:su222maximal}
\mathfrak{u}(1)_R\oplus\mathfrak{su}(2)_{1}\oplus\mathfrak{su}(2)_2\oplus\mathfrak{su}(2)_R\oplus\mathfrak{u}(1)_r\subset\mathfrak{su}(2,2|2)\,.
\end{equation} 
There are essentially three different choices of such subalgebras. Throughout this paper we will require only one and we choose it to contain $\mathcal{Q}_{\alpha}^{I=1,2}$ and $\widetilde{\mathcal{Q}}_{\dot\alpha I=1,2}$ as the $\mathcal{N}=2$ supercharges. This corresponds to $\mathfrak{su}(3)\oplus\mathfrak{u}(1)_{r_{\mathcal{N}=3}}\to\mathfrak{su}(2)_R\oplus\mathfrak{u}(1)_r\oplus\mathfrak{u}(1)_f$. The Cartan of $\mathfrak{su}(2)_R$ is given by $R$ and we take\footnote{Our conventions for $r,R,f$ are chosen to match those of \cite{Nishinaka:2016hbw}.}
\begin{equation}\label{eqn:N2subalgebra}
r=\frac{R_1}{2}+R_2+\frac{R_3}{2}\,,\quad R=\frac{R_1}{2}\,,\quad f=R_3\,.
\end{equation}
Let us now list the branching of the multiplets $\hat{\mathcal{B}}_{[R_1,R_2]}$ under $\mathfrak{su}(2,2|3)\color{black}\to\mathfrak{su}(2,2|2)\oplus\mathfrak{u}(1)_f$. For $\mathfrak{su}(2,2|2)$ multiplets we use the notation of \cite{DolanOsborn}. See also \cite{Cordova:2016emh,Lemos:2016xke} for more general $\mathcal{N}=3\to\mathcal{N}=2$ multiplet decompositions. We have, valid for $R_1R_2\neq0$,
\begin{equation}\label{eqn:Bnmdecomp}
\begin{aligned}
\hat{\mathcal{B}}_{[R_1,R_2]}\color{black}\simeq\,& \hat{\mathcal{B}}^{(R_2-R_1)}_{\frac{R_1+R_2}{2}}\oplus\mathcal{D}^{(R_2-R_1-1)}_{\frac{R_1+R_2-1}{2}(0,0)}\oplus\overline{\mathcal{D}}^{(R_2-R_1+1)}_{\frac{R_1+R_2-1}{2}(0,0)}\oplus\hat{\mathcal{C}}^{(R_2-R_1)}_{\frac{R_1+R_2-2}{2}(0,0)}\\
&\oplus\bigoplus_{i=0}^{R_2-2}\left(\mathcal{B}_{\frac{R_1+i}{2},R_2-i(0,0)}^{(i-R_1)}\oplus\mathcal{C}_{\frac{R_1-1+i}{2},R_2-i-1(0,0)}^{(i-R_1+1)}\right)\\
&\oplus\bigoplus_{i=0}^{R_1-2}\left(\overline{\mathcal{B}}_{\frac{R_2+i}{2},i-R_1(0,0)}^{(R_2-i)}\oplus\overline{\mathcal{C}}_{\frac{R_2+i-1}{2},i+1-R_1(0,0)}^{(R_2-i-1)}\right) \, ,
\end{aligned}
\end{equation}
here the superscript lists the $\mathfrak{u}(1)_f$ charge. Moreover, the above is written with the understanding that any multiplet labelled with a negative value of $R$ is set to zero. The stress-tensor is contained in \eqref{eqn:Bnmdecomp} for $R_1=R_2=1$. We also stress that the $\simeq$ symbol means that the decomposition \eqref{eqn:Bnmdecomp} holds only modulo long multiplets which begin to appear in the decomposition for $R_1R_2\geq4$. For $R_2=0$ the decomposition is
\begin{equation}\label{eqn:Bn0}
\hat{\mathcal{B}}_{[R_1,0]}\color{black}\iso\hat{\mathcal{B}}^{(-R_1)}_{\frac{R_1}{2}}\oplus\overline{\mathcal{D}}^{(1-R_1)}_{\frac{R_1-1}{2}(0,0)}\oplus \overline{\mathcal{E}}_{-R_1(0,0)}^{(0)}\oplus_{i=1}^{R_1-2}\overline{\mathcal{B}}^{(i-R_1+1)}_{\frac{R_1-i-1}{2},-i-1,(0,0)}\,,
\end{equation}
while its conjugate with $R_1=0$ is given by
\begin{equation}\label{eqn:B0n}
\hat{\mathcal{B}}_{[0,R_2]}\color{black}\iso\hat{\mathcal{B}}^{(R_2)}_{\frac{R_2}{2}}\oplus\mathcal{D}_{\frac{R_2-1}{2}(0,0)}^{(R_2-1)}\oplus \mathcal{E}^{(0)}_{R_2(0,0)}\oplus_{i=1}^{R_2-2}\mathcal{B}^{(R_2-i-1)}_{\frac{R_2-i-1}{2},i+1,(0,0)}\,,
\end{equation}
and contains $\mathcal{N}=2$ Coulomb branch operators. We stress that here we use the symbol $\iso$ to indicate that the decompositions \eqref{eqn:Bn0} and \eqref{eqn:B0n} are exact. It is interesting to note that, simply by examining \eqref{eqn:Bnmdecomp}-\eqref{eqn:B0n}, we realize that once we know the Higgs branch ($\mathcal{\hat{B}}_R$ multiplets) we can predict the Coulomb branch ($\mathcal{E}_{r,(0,0)}$ multiplets) but not vice-versa.
\begin{comment}
For completeness we also list the branching that we used for
\begin{equation}
\mathfrak{su}(3)\oplus\mathfrak{u}(1)_{r_{\mathcal{N}=3}}\to\mathfrak{su}(2)_R\oplus\mathfrak{u}(1)_{r}\oplus\mathfrak{u}(1)_{f}\,,
\end{equation}
where
\begin{equation}
\begin{aligned}
\mathbf{[R_1,R_2]}_{\frac{2R_2-2R_1}{3}}\to&\bigoplus_{j=0}^{R_2-1}\bigoplus_{i=0}^{j}\left(\mathbf{\frac{j}{2}}\right)_{R_2-R_1+j-2i}^{2i-j}\oplus\bigoplus_{j=R_2}^{R_1}\bigoplus_{i=0}^{R_2}\left(\mathbf{\frac{j}{2}}\right)_{R_2-R_1+j-2i}^{2i-j}\\
&\oplus\bigoplus_{j=R_1+1}^{R_1+R_2}\bigoplus_{i=0}^{R_2+R_1-j}\left(\mathbf{\frac{j}{2}}\right)_{R_2+R_1-j-2i}^{2i-2R_1+j}\,,
\end{aligned}
\end{equation}
here the notation is $\left(\mathbf{R}\right)_r^f$.
\end{comment}
Note that, as a check, our above syntheses and decompositions in terms of $\mathfrak{su}(2,2|3)$ representations are compatible with the decomposition \cite{DolanOsborn}:
\begin{equation}
\begin{aligned}
\mathcal{B}_{[0,R_2,0]}^{\frac{1}{2},\frac{1}{2}}\color{black}\iso\,&(R_2+1)\hat{\mathcal{B}}_{\frac{R_2}{2}}\oplus\mathcal{E}_{R_2,(0,0)}\oplus\overline{\mathcal{E}}_{-R_2,(0,0)}+(R_2-1)\hat{\mathcal{C}}_{\frac{R_2-2}{2},(0,0)}\oplus R_2\mathcal{D}_{\frac{R_2-1}{2},(0,0)}\\
&\oplus R_2\overline{\mathcal{D}}_{\frac{R_2-1}{2},(0,0)}\oplus\bigoplus_{i=1}^{R_2-2}(i+1)\left(\mathcal{B}_{\frac{i}{2},R_2-i,(0,0)}\oplus\overline{\mathcal{B}}_{\frac{i}{2},i-R_2,(0,0)}\right)\\
&\oplus\bigoplus_{i=0}^{R_2-3}(i+1)\left(\mathcal{C}_{\frac{i}{2},R_2-i-2,(0,0)}\oplus\overline{\mathcal{C}}_{\frac{i}{2},i-R_2+2,(0,0)}\right)\\
&\oplus\bigoplus_{i=0}^{R_2-4}\bigoplus_{j=0}^{R_2-i-4}(i+1)\mathcal{A}^{R_2}_{\frac{i}{2},R_2-i-4-2j,(0,0)}\,.
\end{aligned}
\end{equation}

\section{Indices and the discrete gauging prescription}
\label{sec:Index}
\renewcommand{\arraystretch}{1.1}
Let us introduce the various quantities that we plan to discuss in this paper. 
\subsection{The Superconformal Index}
The superconformal index for $\mathcal{N}=4$ SYM is defined as \cite{Kinney:2005ej,Romelsberger:2005eg}
\begin{equation}
\begin{aligned}\label{eqn:SCI}
 \mathcal{I}^{\mathfrak{g}}\left(t,y,p,q\right)=&\textrm{Tr}_{\mathbb{S}^3}\left[(-1)^Ft^{2(E+j_1)}y^{2j_2}p^{R_2}q^{R_2+2R_3}\right]\\
 =&\textrm{Tr}_{\mathbb{S}^3}\left[(-1)^Ft^{2(E+j_1)}y^{2j_2}(pq)^{r-R}\left(\frac{q^3}{p}\right)^{\frac{f}{2}}\right] \,,
\end{aligned}
\end{equation}
in the second line, since we often wish to treat $\mathcal{N}=4$ SYM as an $\mathcal{N}=2$ theory, we used \eqref{eqn:N2subalgebra} to write the generators in $\mathcal{N}=2$ language. The trace is taken over the Hilbert space of $\mathcal{N}=4$ SYM with gauge algebra $Lie(G)=\mathfrak{g}$ in the radial quantisation. The index \eqref{eqn:SCI} receives contributions only from those states satisfying
\begin{equation}\label{eqn:shortening}
\delta_-^1:=2\left\{\mathcal{Q}_-^{I=1},\mathcal{S}^-_{I=1}\right\}=E-2j_1-\frac{1}{2}(3R_1+2R_2+R_3)=E-2j_1-2R-r=0\,.
\end{equation}
The superconformal index is independent under continuous deformation of the corresponding QFT. In particular
\begin{equation}\label{eqn:couplingind}
\frac{\partial}{\partial \tau} \mathcal{I}^{\mathfrak{g}}\left(t,y,p,q\right)=0\,,
\end{equation}
that is to say \eqref{eqn:SCI} is independent of the gauge coupling $\tau$ of $\mathcal{N}=4$ SYM.
Following \eqref{eqn:couplingind} the superconformal index \eqref{eqn:SCI} may be computed in the free theory by enumerating all of the components of the $\mathcal{N}=4$ field strength multiplet that obey \eqref{eqn:shortening} and then projecting onto gauge invariants. The projection onto gauge invariants is implemented by integration over the gauge group $G$. The index \eqref{eqn:SCI} then takes the form
\begin{equation}\label{eqn:SCIPrescription} 
\mathcal{I}^{\mathfrak{g}}(t,y,p,q)=\int d\mu_{G}(\mathbf{z}) \PE\left[i(t,y,p,q)\chi_{\text{adj}}^{G}(\mathbf{z})\right] \, ,
\end{equation}
$d\mu_{G}$ denotes the Haar measure of the gauge group $G$ and $\chi_{\text{adj}}^{G}$ the character of its adjoint representation. Finally, $\PE\left[f(x)\right]$ denotes the Plethystic exponential of a function $f(x)$, such that $f(0)=0$, given by
\begin{equation}
\PE\left[f(x)\right]:=\exp\left({\sum_{m=1}^{\infty}\frac{1}{m}f(x^m)}\right)\,.
\end{equation}
\begin{table}
\centering
\begin{tabular}{|c||c|c|c|c|c|c|c|} 
\hline
\textrm{Letters}  & $E$ & $j_1$ & $j_2$ & $R_1$ & $R_2$ & $R_3$ & $i(t,y,p,q,\epsilon)$\\ 
  \hline\hline
 $F_{++}$ & $2$ & $1$ & $0$ & $0$ & $0$ & $0$ &$t^6$\\ 
 \hline
   $\bar{\lambda}^{I=1}_{\dot{\pm}}$ & $\frac{3}{2}$ & $0$ & $\pm\frac{1}{2}$ & $1$ & $0$ & $0$&$- t^3(y+y^{-1})$\\ 
 \hline
   $\lambda_{-I=2,3,4}$ & $\frac{3}{2}$ & $\frac{1}{2}$ & $0$ & $1,0,0$ & $-1,1,0$ & $0,-1,1$& $-t^4\left(\frac{1}{ pq}+\frac{p}{ q}+ q^2\right)$\\ 
 \hline
    $X,Y,Z$ & $1$ & $0$ & $0$ & $0,1,1$ & $1,-1,0$ & $0,1,-1$ & $t^2\left( pq+\frac{ q}{p}+\frac{1}{ q^2}\right)$\\ 
 \hline\hline
      $\partial\bar{\lambda}^1=0$& $\frac{5}{2}$ & $\frac{1}{2}$ & $0$ & $1$ & $0$ & $0$ &$ t^6$\\ 
 \hline\hline
   $\partial_{+\dot\pm}$ & $1$ & $\frac{1}{2}$ & $\pm\frac{1}{2}$ & $0$ & $0$ & $0$ &$t^3y$ , $t^3y^{-1}$\\ 
 \hline
\end{tabular}
\caption{The on-shell degrees of freedom of the $\mathcal{N}=4$ field strength multiplet are $F_{\alpha\beta}$, $\widetilde{F}_{\dot\alpha\dot\beta}$, $\lambda_{\alpha I}$, $\bar{\lambda}_{\dot\alpha}^I$, $\phi^{IJ}$ with $I=1,2,3,4$ and $\phi^{IJ}$ is in the $\mathbf{[0,1,0]}$ of $\mathfrak{su}(4)$. We define $X=\phi^{12}$, $Y=\phi^{13}$ and $Z=\phi^{14}$. $\partial\bar{\lambda}^1$ denotes the equation of motion $\partial_{+\dot+}\bar{\lambda}_{\dot{-}}^1+\partial_{+\dot-}\bar{\lambda}^1_{\dot{+}}=0$ which enters with opposite statistics.
% Knowledge of the $\mathcal{N}=3$ structure \eqref{eqn:ik} allows us to write down the charges under $r_{\kgg}\cdot s_{\kgg}$ for each letter
\label{fig:N4vecletters}}
\end{table}
The single letter index $i(t,y,p,q)$ may be computed by enumerating all letters with $\delta_-^1=0$, listed in Table \ref{fig:N4vecletters}. Or equivalently by evaluating the index of the $\mathfrak{psu}(2,2|4)$ multiplet $\mathcal{B}_{[0,1,0]}^{\frac{1}{2},\frac{1}{2}}$, which is the free $\mathcal{N}=4$ vector multiplet plus conformal descendents. It is given by
\begin{equation}\label{eqn:singleletter}
i(t,y,p,q)=\mathcal{I}_{\mathcal{B}_{[0,1,0]}^{\frac{1}{2},\frac{1}{2}}}=\frac{\left(p^{-1}q+pq+q^{-2}\right)t^2-\chi_1(y)t^3-\left(q^2+p^{-1}q^{-1}+pq^{-1}\right)t^4+2t^6}{(1-t^3y)(1-t^3y^{-1})} \, ,
\end{equation}
where $\chi_{2j_2}(y)$ denotes the $SU(2)$ character given by
\begin{equation}
\chi_s(y)\equiv\chi_s=y^{s}+y^{s-2}+\dots+y^{-s}\,.
\end{equation}
The index \eqref{eqn:SCI} counts short representations of the $\mathfrak{su}(2,2|4)$ superconformal algebra, modulo recombination.
Meaning that all short multiplets, see \eqref{eq:MultiIndexE}-\eqref{eqn:index0}, contribute to the index, however, when they satisfy the recombination rules \eqref{eqn:short1}-\eqref{eqn:short2} they sum to zero.
Recombination happens when a long multiplet $\mathcal{A}^E_{[R_1,R_2,R_3],(j_1,j_2)}$ hits the unitary bound and decomposes into semi-direct sums of short representations. We list the possible recombination rules, viewing as an $\mathcal{N}=2$ theory, in equations \eqref{eqn:short1}-\eqref{eqn:short2}. The index \eqref{eqn:SCI} can therefore be expanded in the following form
\begin{equation}
\mathcal{I}^{\mathfrak{g}}(t,y,p,q) =
\sum_{\mathcal{M}_{\mathcal{N}=4} \in \text{shorts}} \mathcal{I}_{\mathcal{M}_{\mathcal{N}=4}}(t,y,p,q) \, ,
\end{equation}
where the sum is taken over the short multiplets of the theory, modulo those that can recombine into long multiplets.
We list the indices of multiplets of an $\mathfrak{su}(2,2|2)\subset \mathfrak{psu}(2,2|4)$ subalgebra in Appendix \ref{sec:indicies}. 
As we discussed in Section \ref{sec:N3Cons} at $\tau=e^{\pi\iu/3},\iu,e^{\pi\iu/3}$ the global symmetry group (at the level of local operators) of the theory has a $\mathbb{Z}_{\kgg}$ enhancement. Correspondingly the Hilbert space has an extra $\mathbb{Z}_{\kgg}$ grading at those values of the coupling. Therefore one may define a further refined version of the superconformal index given by
\begin{equation}\label{eqn:SCIrefined}
 \mathcal{I}^{\mathfrak{g}}\left(t,y,p,q,\epsilon\right)
 =\textrm{Tr}_{\mathbb{S}^3}\left[(-1)^Ft^{2(E+j_1)}y^{2j_2}(pq)^{r-R}\left(\frac{q^3}{p}\right)^{\frac{f}{2}} \epsilon^{r_{\kgg}\newdot s_{\kgg}}\right]\, ,
\end{equation}
where we introduced the $\mathbb{Z}_{\kgg}$-valued fugacity $\epsilon$ in order to keep track of the discrete symmetry. We stress that the $\mathbb{Z}_{\kgg}$ is a global symmetry only at $\tau=e^{\pi\iu/3},\iu,e^{\pi\iu/3}$. As we showed in Appendix \ref{sec:preservedSCA} the $\mathbb{Z}_{\kgg}$ commutes with the supercharges $\mathcal{Q}_{-}^{I=1}$ and $\mathcal{S}^{-}_{I=1}$ that we used to compute the index \eqref{eqn:SCI} with respect to. Moreover, we also demonstrated that the $\mathbb{Z}_{\kgg}$ preserves a $\mathfrak{su}(2,2|3)\subset\mathfrak{psu}(2,2|4)$ subalgebra and it therefore preserves the recombination rules \eqref{eqn:short1}-\eqref{eqn:short2}.

Therefore the refined index \eqref{eqn:SCIrefined} can again be expanded
\begin{equation}\label{eqn:epsilonpres}
\begin{aligned}
\mathcal{I}^{\mathfrak{g}}(t,y,p,q,\epsilon) =&
\sum_{\mathcal{M}_{\mathcal{N}=4} \in \text{shorts}} \mathcal{I}_{\mathcal{M}_{\mathcal{N}=4}}(t,y,p,q,\epsilon)\\
=&\sum_{\bigoplus_i\mathcal{M}^{(i)}_{\mathcal{N}=3} \in \text{shorts}} \epsilon^{r_{\kgg}\left(\mathcal{M}_{\mathcal{N}=3}\right)+ s_{\kgg}\left(\mathcal{M}_{\mathcal{N}=3}\right)}\mathcal{I}_{\mathcal{M}_{\mathcal{N}=3}}(t,y,p,q)\,.
\end{aligned}
\end{equation}
In the final equality, we firstly used the fact that any short multiplet $\mathcal{M}_{\mathcal{N}=4}$ of $\mathfrak{psu}(2,2|4)$ can be decomposed into multiplets of a $\mathfrak{su}(2,2|3)$ subalgebra $\mathcal{M}_{\mathcal{N}=4}\iso\bigoplus_i\mathcal{M}^{(i)}_{\mathcal{N}=3}$. Secondly we used the fact that the action of $r_{\kgg}\newdot s_{\kgg}$ preserves the $\mathfrak{su}(2,2|3)\subset\mathfrak{psu}(2,2|4)$ subalgebra and by $r_{\kgg}(\mathcal{M}_{\mathcal{N}=3})\newdot s_{\kgg}(\mathcal{M}_{\mathcal{N}=3})$ we mean the generator of $\mathbb{Z}_{\kgg}$ evaluated on the given multiplet. For example, using \eqref{B0m0}, the refined index on the free $\mathcal{N}=4$ vector multiplet is given by
\begin{equation}\label{eqn:ik}
\begin{aligned}
&\mathcal{I}_{\mathcal{B}_{[0,1,0]}^{\frac{1}{2},\frac{1}{2}}}(t,y,p,q,\epsilon)=\epsilon^{-1}\,\mathcal{I}_{\hat{\mathcal{B}}_{[1,0]}}(t,y,p,q)+\epsilon\,\mathcal{I}_{\hat{\mathcal{B}}_{[0,1]}}(t,y,p,q)\\
&=\epsilon^{-1}\frac{q^{-2}t^2-\left(p^{-1}q^{-1}+pq^{-1}\right)t^4+t^6}{(1-t^3y)(1-t^3y^{-1})}+\epsilon\frac{p^{-1}qt^2+pqt^2-\chi_1(y)t^3-q^2t^4+t^6}{(1-t^3y)(1-t^3y^{-1})}
\end{aligned}
\end{equation}
We may then gauge the discrete $\mathbb{Z}_{\kgg}$ symmetry by making the projection
\begin{equation}\label{eqn:discgaugepro}
\mathcal{I}_{\mathbb{Z}_{\kgg}}^{\mathfrak{g}}(t,y,p,q):=\frac{1}{|\mathbb{Z}_{\kgg}|}\sum_{\epsilon\in\mathbb{Z}_{\kgg}}\mathcal{I}^{\mathfrak{g}}(t,y,p,q,\epsilon) \, .
\end{equation} 
The discrete gauging restricts each contribution, in terms of either $\mathfrak{su}(2,2|3)$ or $\mathfrak{su}(2,2|2)$ multiplets, to satisfy
\begin{equation}\label{eqn:HLCond}
r_{\kgg}\newdot s_{\kgg}=r+f\newdot s_{\kgg}=0\bmod {\kgg}\,.
\end{equation}
We demonstrate in Section \ref{sec:largeN} that \eqref{eqn:epsilonpres} reproduces the {\it refined} superconformal index \eqref{eqn:SCIrefined} at large $N$ by matching with the KK supergraviton index (\ref{eqn:LargeN}) obtained by the AdS/CFT computation of \cite{Imamura:2016abe}. We would like stress that the final expression for KK supergraviton index is {\it not equal} to the index of a theory obtained through an S-fold projection. Since, as shown in Section \ref{sec:largeN}, this last expression is obtained implementing the orbifold projection at the level of the single particle index.

\subsection{Coulomb branch limit}
The graded index \eqref{eqn:SCIrefined} may be rewritten as \cite{Gadde:2011uv}
\begin{equation}\label{eqn:SCISchur}
\mathcal{I}^{\mathfrak{g}}\left(t,y,p,q,\epsilon\right)= \textrm{Tr}_{\mathbb{S}^3}\left[(-1)^F\tau^{\frac{1}{2}\delta_+^2}\sigma^{\frac{1}{2}\widetilde{\delta}_{\dot+ 2}}\rho^{\frac{1}{2}\widetilde{\delta}_{\dot- 2}}u_f^f \epsilon^{r_{\kgg}\newdot s_{\kgg}}\right]\, ,
\end{equation}
with
\begin{equation}
\tau:=\frac{t^2}{\sqrt{pq}}\,,\quad \sigma:=ty\sqrt{pq}\,,\quad \rho:=\frac{t\sqrt{pq}}{y}\,,\quad u_f:=\sqrt{\frac{q^3}{p}}\,,
\end{equation}
and
\begin{align}
&\delta_{\pm}^2:=2\left\{\mathcal{Q}_{\pm}^2,\left(\mathcal{Q}_{\pm}^2\right)^{\dagger}\right\}=E\pm2j_1+2R-r\,,\\
&\widetilde{\delta}_{\dot\pm 2}:=2\left\{\widetilde{\mathcal{Q}}_{\dot\pm 2},\left(\widetilde{\mathcal{Q}}_{\dot\pm 2}\right)^{\dagger}\right\}=E\pm2j_2-2R+r\,.
\end{align}
In the parametrisation \eqref{eqn:SCISchur} the Coulomb branch limit of the superconformal index is defined to be \cite{Gadde:2011uv}
\begin{equation}\label{eqn:CBlimit}
\tau\to0\,,\quad \rho\,,\sigma\,\text{ fixed,}
\end{equation}
which is well defined since $\delta_+^2\geq0$. In this limit the index is then given by
\begin{equation}\label{eqn:SCICB1}
\mathcal{I}^{\mathfrak{g}}_{\text{CB}}\left(\rho,\sigma,u_f,\epsilon\right)= \textrm{Tr}_{\mathbb{S}^3|\delta_+^2=0}\left[(-1)^F\sigma^{\frac{1}{2}\widetilde{\delta}_{\dot+ 2}}\rho^{\frac{1}{2}\widetilde{\delta}_{\dot- 2}}u_f^f\epsilon^{r_{\kgg}\newdot s_{\kgg}}\right]\,.
\end{equation}
Defining
\begin{equation}
\rho\sigma=x\,,\quad \rho/\sigma=v \, ,
\end{equation}
the single letter index \eqref{eqn:singleletter} in the Coulomb branch limit becomes
\begin{equation}
i_{\text{CB}}(x)=x\,.
\end{equation}
In our $\mathcal{N}=2$ decomposition this is simply the contribution of the single letter $X$ described in Table \ref{fig:N4vecletters}. 
Since, for our theories, it is independent of both the ratio $v=\rho/\sigma$ and $u_f$ then, due to $\left(\widetilde{\delta}_{\dot+2}+\widetilde{\delta}_{\dot-2}\right)\mathcal{Q}_{\alpha}^1=\left(\widetilde{\delta}_{\dot+2}+\widetilde{\delta}_{\dot-2}\right)\mathcal{Q}^2_{+}=0$, \eqref{eqn:SCICB1} is further shortened and preserves $\mathcal{Q}^1_{\pm}$, $\mathcal{Q}^2_{+}$. This allows us to write 
\begin{equation}
E=r\,,\quad j_1=j_2=f=R=0\,,
\end{equation}
these are the highest weight states of the $\mathcal{N}=2$ $\mathcal{E}_{r,(0,0)}$ multiplets that generate the Coulomb branch chiral ring. Therefore the only non-zero contributions to \eqref{eqn:SCICB1} are from 
\begin{equation}
\mathcal{I}_{\mathcal{E}_{r,(0,0)}}(x,\epsilon=1)=x^r\,.
\end{equation} 
Hence, following \eqref{eqn:disgaugeZnaction}, the prescription \eqref{eqn:epsilonpres} can be implemented simply by
\begin{equation}
x\to\epsilon x \, ,
\end{equation}
and the index can be written as
\begin{equation}
\label{eq:indexCoulomb}
\mathcal{I}^{\mathfrak{g}}_{\text{CB}}\left(x,\epsilon\right)= \textrm{Tr}_{\mathbb{S}^3|\delta_+^2=0}\left[\epsilon^rx^{r}\right]= \int d\mu_{G}(\mathbf{z}) \PE\left[ i_{\text{CB}}(\epsilon x)\,\chi_{\text{adj}}^{G}(\mathbf{z})\right] \,.
\end{equation}
We would like to stress that the $\mathcal{E}_{r,(0,0)}$ multiplets do not recombine \cite{DolanOsborn} and therefore turning on the refinement $\epsilon$ for the discrete symmetry commutes with the integration over $G$. 
Let $G$ be connected then, as pointed out in \cite{Gadde:2011uv},  (\ref{eq:indexCoulomb}) may be explictly evaluated thanks to Macdonald's constant-term identities \cite{macdonald1998symmetric,macdonald}
\begin{equation}\label{eqn:SCICBdeg}
\mathcal{I}^{\mathfrak{g}}_{\text{CB}}\left(x,\epsilon\right)= \PE\left[\sum_{j\in \exponents(\mathfrak{g})}\epsilon^{j+1}x^{j+1}\right]\,,
\end{equation}
where $\exponents(\mathfrak{g})$ denotes the set of exponents of the Lie algebra $\mathfrak{g}=Lie(G)$. The elements of $\exponents(\mathfrak{g})$ are in one-to-one correspondence with the degrees of the generators of the ring of $\mathfrak{g}$-invariant polynomials. We list the elements of $\exponents(\mathfrak{g})$ for $\mathfrak{g}=ADE$ and $\mathfrak{u}(N)$ in Table \ref{tab:degrees}.
\begin{table}
\centering
\begin{tabular}{|c|c|} 
\hline
 $\mathfrak{g}$ & $\exponents(\mathfrak{g})$\\\hline 
 \hline
  $\mathfrak{u}(N)$&$0,1,2,\dots,N-1$\\\hline
 $A_{N}$&$1$, $2$, $3$, $\dots$, $N$\\\hline
% $B_{N}$&$1$, $3$, $5$, $\dots$, $2N-1$\\\hline
% $C_{N}$&$1$, $3$, $5$, $\dots$, $2N-1$\\\hline
 $D_{N}$&$1$, $3$, $5$, $\dots$, $2N-3$; $N-1$\\\hline
 $E_{6}$&$1$, $4$, $5$, $7$, $8$, $11$\\\hline
 $E_{7}$&$1$, $5$, $7$, $9$, $11$, $13$, $17$\\\hline
 $E_{8}$&$1$, $7$, $11$, $13$, $17$, $19$, $23$, $29$\\\hline
% $F_{4}$&$1$, $5$, $7$, $11$\\\hline
% $G_{2}$&$1$, $5$\\\hline
\end{tabular}
\caption{Exponents of the Lie algebra $\mathfrak{g}$.}
\label{tab:degrees}
\end{table}
According to \eqref{eqn:discgaugepro}, upon the discrete gauging, we then have
\begin{equation}\label{eqn:SCICB}
\mathcal{I}^{\mathfrak{g}}_{\mathbb{Z}_{\kgg},\text{CB}}\left(x\right)= \frac{1}{|\mathbb{Z}_{\kgg}|}\sum_{\epsilon\in\mathbb{Z}_{\kgg}}\mathcal{I}^{\mathfrak{g}}_{\text{CB}}\left(x,\epsilon\right) \,.
\end{equation}
Since \eqref{eqn:SCICB} counts only gauge invariant chiral operators, it is equal to the Coulomb branch Hilbert series for the discretely gauged theory. Therefore the rank, i.e. the complex dimension of the Coulomb branch $CB_{\mathfrak{g},n}$, is equal to \cite{Cremonesi:2017jrk}
\begin{equation}\label{eqn:CBdim}
\dim_{\mathbb{C}}CB_{\mathfrak{g},n}=\left(\text{Order of pole at $x=1$ of $\mathcal{I}_{\mathbb{Z}_{\kgg},\text{CB}}^{\mathfrak{g}}(x)$}\right)\,.
\end{equation} 
We of course expect that $\dim_{\mathbb{C}}CB_{\mathfrak{g},n}=\rank\mathfrak{g}$. In the following sections we analyse some examples. 

\subsection{Higgs branch Hilbert series}
In general the Hilbert series \cite{Benvenuti:2006qr,Feng:2007ur} counts gauge invariant chiral operators graded by their charges under a maximally commuting subalgebra of the global symmetry algebra. We will be interested in computing the Hilbert series for the Higgs branch $HB_{\mathfrak{g}}$ of $\mathcal{N}=4$ SYM (using the $\mathcal{N}=2$ decomposition \eqref{eqn:N2subalgebra}). This is given by
\begin{equation}\label{eqn:hs}
\HS^{\mathfrak{g}}(\tHS,u_f,\epsilon): = \Tr_{\mathcal{H}}\left[\tHS^{2R}u_f^f\epsilon^{r_{\kgg}\newdot s_{\kgg}}\right] \, ,
\end{equation}
where $\mathcal{H}=\{\mathcal{O}_i | \widetilde{\mathcal{Q}}^{I}_{\dot{\alpha}}\mathcal{O}_i=0\,, M_{\mu\nu} \mathcal{O}_i=0\,,r\mathcal{O}_i =0 \}$ is the space of scalar, $\mathfrak{g}$-invariant chiral operators that parametrize the Higgs branch moduli space of vacua. In the language of the previous section \eqref{eqn:hs} is counting $\hat{\mathcal{B}}_R$ operators with $E=2R$ and $r=j_1=j_2=0$. We stress that there is no recombination rule \eqref{eqn:short1}-\eqref{eqn:short2} involving only $\hat{\mathcal{B}}_R$ operators.

In the $\mathcal{N}=2$ decomposition that we used in Section \ref{sec:su22Nrepn}, the Higgs branch for our theories is reached by setting equal to zero the scalar field $X$ in the $\mathcal{N}=2$ vector multiplet. Therefore there is only one relevant F-term that we must take into account
\begin{equation}\label{eqn:Fsuper}
\partial_{X}\mathcal{W} = [Y,Z] = 0 \, ,
\end{equation}
where $\mathcal{W}$ is the superpotential for $\mathcal{N}=4$ SYM. Unfortunately, due to the fact that the gauge group is not completely broken, letter counting techniques cannot be used to compute \eqref{eqn:hs}. Instead, in order to compute \eqref{eqn:hs}, we use the package \textit{Macaulay2} \cite{M2}. By inputting the ring of polynomials $\mathcal{R}=\mathbb{C}[Y,Z]$ and the ideal $I$ given by \eqref{eqn:Fsuper}, \textit{Macaulay2} can compute the Hilbert series for $\mathcal{R}/I$.

Since both $r$ and $s_{\kgg}$ act trivially on the fields $Y,Z$; on the Higgs branch $r_{\kgg}\newdot s_{\kgg}=f$. Therefore the extra grading may be implemented by $u_f\to\epsilon u_f$. The Higgs branch Hilbert series then takes the form
\begin{equation}
\HS^{\mathfrak{g}}(\tHS,u_f,\epsilon)=\int d\mu_{G}(\mathbf{z})\mathcal{F}^{\flat}_{\kgg}(\tHS,\epsilon u_f,\mathbf{z})\,,
\end{equation}
where $\mathcal{F}^{\flat}_{\kgg}(\tHS,u_f,\mathbf{z})$ denotes the F-flat Hilbert series for $\mathcal{N}=4$ SYM. The discrete gauged Higgs branch Hilbert series reads
\begin{equation}
\label{eq:HIggsHS}
\HS^{\mathfrak{g}}_{\mathbb{Z}_{\kgg}}(\tHS,u_f)=\frac{1}{|\mathbb{Z}_{\kgg}|}\sum_{\epsilon\in\mathbb{Z}_{\kgg}}\HS^{\mathfrak{g}}(\tHS,u_f,\epsilon)\, .
\end{equation}
One important piece of information carried by \eqref{eqn:hs} is the dimension of the Higgs branch 
\begin{equation}\label{eqn:dimHB}
\dim_{\mathbb{C}}HB_{\mathfrak{g},n}=\left(\text{Order of pole at $\tHS=1$ of $\HS^{\mathfrak{g}}_{\mathbb{Z}_{\kgg}}(\tHS,1)$}\right)\,.
\end{equation} 
A particularly useful quantity is the Plethystic logarithm of the Hilbert series $\PLog\left[\HS^{\mathfrak{g}}_{\mathbb{Z}_{\kgg}}\right]$. The Plethystic logarithm is defined as
\begin{equation}\label{eqn:PLog}
\PE^{-1}\left[f(x)\right]=\PLog\left[f(x)\right]:=\sum_{m=1}^{\infty}\frac{\mu(n)}{m}\log\left(f(x^m)\right)\,,
\end{equation}
where $\mu(m)$ is the M\"{o}bius $\mu$ function.
The Plethystic logarithm of the Hilbert series satisfies \cite{Benvenuti:2006qr,DelZotto:2014kka}:
\begin{itemize}
\item When the moduli space is a complete intersection variety $\PLog\left[\HS^{\mathfrak{g}}_{\mathbb{Z}_{\kgg}}(\tHS,u_f)\right]$ is a polynomial of finite degree. When it is not the $\PLog\left[\HS^{\mathfrak{g}}_{\mathbb{Z}_{\kgg}}\right]$ is an infinite series in $\tHS$.
\item It has been conjectured in \cite{Feng:2007ur,Benvenuti:2006qr} that when the moduli space is a complete intersection variety the first coefficients with positive sign in the $\PLog\left[\HS^{\mathfrak{g}}_{\mathbb{Z}_{\kgg}}\right]$ polynomial encode the generators of the variety. Negative coefficients encode relations. When the moduli space is not a complete intersection the generators of the moduli space are generally still captured by the first positive terms. However, in this last case, most of the contributions in the PLog expansion are redundant and represent Hilbert syzygies.
\end{itemize}
Of course this discussion also applies to the Coulomb branch index \eqref{eqn:SCICB}.

\section{Rank 1 theories}
\label{sec:rank1}
Having introduced the main quantities that we wish to compute we will now go ahead and compute them for the possible $\mathcal{N}=3$ rank one theories that can be obtained via discrete gauging of $\mathcal{N}=4$ SYM. As we mentioned previously, if we restrict to connected groups then, from the point of view of the superconformal index there are only two distinct possibilities, labelled by the two choices of Lie algebras of rank one i.e. $\mathfrak{g}=\mathfrak{u}(1)$ and $\mathfrak{g}=\mathfrak{su}(2)$.
\subsection{$\mathfrak{g}=\mathfrak{u}(1)$}
Let us begin with the $\mathbb{Z}_n$ gauging of $\mathfrak{g}=\mathfrak{u}(1)$ $\mathcal{N}=4$ SYM. 
%Before the discrete gauging this theory is actually equivalent to the S-fold theory $S_{k,\ell}^{N}$ with $\ell=N=1$ which has enhanced $\mathcal{N}=4$ supersymmetry \cite{Aharony:2015oyb,Aharony:2016kai,Imamura:2016udl,Agarwal:2016rvx}. This therefore implies that the superconformal index that we will compute will be equal to the superconformal index for the S-fold theory $S_{k,1,p'}^1$ of \eqref{eqn:sfolddisgauge} with $k=p'=\kgg$. 
\subsubsection*{Superconformal index}
%\textcolor{red}{For $\mathfrak{g}=\mathfrak{u}(1)$ the contributions arising from the different $\mathcal{N}=3$ in (\ref{eqn:epsilonpres}) can be summed up using a PE, is this correct?}. Therefore the index for the discrete gauging of the $\mathbb{Z}_{\kgg}$ generated by $r_{\kgg}\cdot s_{\kgg}$ may be easily computed. It reads
%For $\mathfrak{g}=\mathfrak{u}(1)$ the adjoint character is $\chi_{\text{adj}}(z)=1$ and therefore the integration over the gauge group is trivially performed.
Since the $\mathfrak{u}(1)$ $\mathcal{N}=4$ theory is free the index for the discrete gauging can be computed explicitly. Using \eqref{eqn:ik}, it is given by 
\begin{equation}\label{eqn:u1index}
\mathcal{I}^{\mathfrak{u}(1)}_{\mathbb{Z}_{\kgg}}(t,y,p,q)=\frac{1}{|\mathbb{Z}_{\kgg}|}\sum_{\epsilon\in \mathbb{Z}_{\kgg}}\mathcal{I}^{\mathfrak{u}(1)}_{\mathbb{Z}_{\kgg}}(t,y,p,q,\epsilon)=\frac{1}{|\mathbb{Z}_{\kgg}|}\sum_{\epsilon\in \mathbb{Z}_{\kgg}}\PE\left[\epsilon^{-1}\,\mathcal{I}_{\hat{\mathcal{B}}_{[1,0]}\color{black}}+\epsilon\,\mathcal{I}_{\hat{\mathcal{B}}_{[0,1]}\color{black}}\right]\, .
\end{equation}
\begin{comment}
The index may be equivalently expressed in terms of Elliptic Gamma functions  \cite{Dolan:2008qi}
\begin{equation}
\mathcal{I}^{\mathfrak{u}(1)}_{\mathbb{Z}_{\kgg}}(t,y,p,q)=\frac{1}{|\mathbb{Z}_{\kgg}|}\sum_{\epsilon\in\mathbb{Z}_{\kgg}}\frac{\Gamma\left(\frac{\epsilon qt^2}{p};t^3y,\frac{t^3}{y}\right)\Gamma\left(\frac{t^2}{\epsilon q^2};t^3y,\frac{t^3}{y}\right)\Gamma\left(\epsilon pqt^2;t^3y,\frac{t^3}{y}\right)}{\left(\epsilon t^3y;t^3y\right)^{-1}\left(\frac{t^3y}{\epsilon };t^3y\right)^{-1}\Gamma\left(\frac{t^3}{\epsilon y};t^3y,\frac{t^3}{y}\right)}\,,
\label{eqn:u1index}
\end{equation}
where 
\begin{equation}\label{eqn:ElliGamma}
\Gamma(z;w,v):=\prod_{j,m=0}^{\infty}\frac{1-z^{-1}w^{j+1}v^{m+1}}{1-zw^{j}v^{m}}\,,\quad (x;q):=\prod_{j=0}^{\infty}(1-xq^j)\,,
\end{equation}
defines the Elliptic Gamma function and $q$-Pochammer symbol respectively.
\end{comment}
When $\kgg=2$ the expression \eqref{eqn:u1index} is exactly the index for the $G=O(2)$ $\mathcal{N}=4$ theory which matches the expectation that the this theory is nothing but the usual $O3^-$ orientifold theory. 
We can perform several other checks of our expression \eqref{eqn:u1index} by studying the various limits that we outlined in Section \ref{sec:Index}.
\subsubsection*{Coulomb branch limit}
After taking the Coulomb branch limit \eqref{eqn:CBlimit} we find, for the discrete gauging of the $\mathfrak{g}=\mathfrak{u}(1)$ theory,
\begin{equation}\label{eqn:u1CBlimit}
\mathcal{I}^{\mathfrak{u}(1)}_{\mathbb{Z}_{\kgg},\text{CB}}\left(x\right)=\frac{1}{{\kgg}}\sum_{\epsilon\in\mathbb{Z}_{\kgg}}\PE\left[\epsilon x\right]=\PE\left[x^{\kgg}\right]\,.
\end{equation}
This implies that the Coulomb branch is freely generated by $\widetilde{u}=u^{\kgg}$ with $u=X$ the parent Coulomb branch parameter. Therefore $E(\widetilde{u})=r(\widetilde{u})={\kgg}$ which implies that the $\widetilde{u}$ is the superconformal primary of the $\mathfrak{su}(2,2|2)$ multiplet $\mathcal{E}_{{\kgg},(0,0)}\subset \hat{\mathcal{B}}_{[0,{\kgg}]}$. The topology is simply $CB_{\mathfrak{u}(1),\kgg}=\mathbb{C}[\widetilde{u}]\iso\mathbb{C}$. We wish to point out that \eqref{eqn:u1CBlimit} is in perfect agreement with the expected spectrum of Coulomb operators \eqref{eqn:sfolddiscgauge} coming from the S-fold analysis \cite{Aharony:2016kai} and the Seiberg-Witten curve analysis for the quotient of the $I_{0}$ geometry in the discussion below equation (2.8) of \cite{Argyres:2016yzz}.
\subsubsection*{Higgs branch Hilbert series}
We now compute the Higgs branch Hilbert Series for these theories. For $\mathfrak{g}=\mathfrak{u}(1)$ the superpotential \eqref{eqn:Fsuper} is trivial and we may actually use letter counting. We find that $\mathcal{F}^{\flat}_{\kgg}(\tHS,u_f,\epsilon)=\PE\left[\epsilon u_f\tHS+\epsilon^{-1}u_f^{-1}\tHS\right]$. The integration over the gauge group is trivially performed and we get
\begin{equation}
\begin{aligned}
\label{eq:hsu1}
\HS^{\mathfrak{u}(1)}_{\mathbb{Z}_{\kgg}}(\tHS,u_f) &= \frac{1}{|\mathbb{Z}_{\kgg}|}\sum_{\epsilon \in \mathbb{Z}_{\kgg}}\mathcal{F}^{\flat}_{\kgg}(\tHS,\epsilon u_f)=\PE\left[\tHS^2+\chi_1(u_f)\tHS^{\kgg}-\tHS^{2{\kgg}}\right]\\
&=\text{Hilbert Series of $\mathbb{C}^2/\mathbb{Z}_{\kgg}$}\, .
\end{aligned}
\end{equation}
The generators are simply given by 
\begin{equation}
W^+=Y^{\kgg}\,,\quad W^-=Z^{\kgg}\,,\quad J=YZ\,.
\end{equation} 
They satisfy the relation $W^+W^-=J^{\kgg}$. In terms of $\mathfrak{su}(2,2|3)$ multiplets this is equivalently expressed as
\begin{equation}\label{eqn:su223relation}
\hat{\mathcal{B}}_{[{\kgg},0]}\hat{\mathcal{B}}_{[0,{\kgg}]}\color{black}\sim\left(\hat{\mathcal{B}}_{[1,1]}\color{black}\right)^{\kgg}\,.
\end{equation}
The topology of the moduli space and relation are in perfect agreement with equations (2.1) and (2.16), respectively, of \cite{Nishinaka:2016hbw,Lemos:2016xke}.

\subsection{$\mathfrak{g}=\mathfrak{su}(2)$}\label{sec:su2} 
For $\mathfrak{g}=\mathfrak{su}(2)$ it is very difficult to compute \eqref{eqn:SCI} in closed form. For this reason we will instead study only the Coulomb branch limit of the index and the Higgs branch Hilbert series.
\subsubsection*{Coulomb branch limit}
Let us now study the Coulomb branch limit (\ref{eqn:SCICB}). The corresponding computation can be easily performed and we get
\begin{equation}
\mathcal{I}_{\mathbb{Z}_{\kgg},\text{CB}}^{\mathfrak{su}(2)}(x)=\frac{1}{|\mathbb{Z}_{\kgg}|}\sum_{\epsilon\in\mathbb{Z}_{\kgg}}\PE\left[\epsilon^2x^2\right]=\begin{cases}\PE\left[x^2\right]&{\kgg}=1\\
\PE\left[x^{\kgg}\right]&{\kgg}=2,4,6\\
\PE\left[x^6\right]&{\kgg}=3\\\end{cases}\,.
\end{equation}
The topology in each case is $CB_{\mathfrak{su}(2),n}=\mathbb{C}\left[\widetilde{u}\right]\iso\mathbb{C}$. For ${\kgg}=2,4,6$ the Coulomb branch of the discretely gauged theory, $CB_{\mathfrak{su}(2),n}$, is generated by $\widetilde{u}=u^{{\kgg}/E(u)}$ where $u=\frac{1}{2}\tr X^2$ is the Coulomb branch parameter of the parent theory. Therefore $E(\widetilde{u})=r(\widetilde{u})={\kgg}$ which belong to $\mathcal{E}_{{\kgg},(0,0)}\subset\hat{\mathcal{B}}_{[0,{\kgg}]}$ for ${\kgg}=2,4,6$. This matches with the discussion below equation (2.8) of \cite{Argyres:2016yzz} for the $I_4$-series $I_0^*$ geometries. The ${\kgg}=3$ case is slightly different since $E(u)=2$ is not a divisor of ${\kgg}=3$ and $CB_{\mathfrak{su}(2),3}$ is generated by $\widetilde{u}=u^{\kgg}=u^3$. Nevertheless this is in perfect agreement with the discussion below equation (A.7) of \cite{Argyres:2016yzz} for the $I_2$-series $I_0^*$ geometries. These parent theories do not come from S-folds and so do not fall into the considerations of \cite{Aharony:2016kai}.
\subsubsection*{Higgs branch Hilbert series}
Let us now compute the Higgs branch Hilbert series \eqref{eqn:hs}. For the case at hand the gauge group is not completely broken and we cannot use letter counting. Therefore we compute the F-flat Hilbert series using \textit{Macaulay2}. We obtain
\begin{equation}
\label{eq:hssu2}
\mathcal{F}^{\flat}_{\kgg}(\tHS,\epsilon u_f,z)=\left(1-\chi_2(z)\tHS^2+\left(\epsilon u_f+\frac{1}{\epsilon  u_f}\right)\tHS^3\right)\PE\left[\tHS\left(\epsilon u_f+\frac{1}{\epsilon u_f}\right)\chi_2(z)\right] \, .
\end{equation}
Note that the same result was already found, for $\kgg=1$, in \cite{Hanany:2010qu}. After the integration over the $SU(2)$ gauge group we get
\begin{equation}
\begin{aligned}
\label{eq:hssu2}
\HS_{\mathbb{Z}_{\kgg}}^{\mathfrak{su}(2)}(\tHS,u_f)& = \frac{1}{|\mathbb{Z}_{\kgg}|}\sum_{\epsilon \in \mathbb{Z}_{\kgg}}\int d\mu_{SU(2)}(z)\mathcal{F}^{\flat}_{\kgg}(\tHS,\epsilon u_f,z)\\
&= \frac{1}{\kgg}\sum_{\epsilon \in \mathbb{Z}_{\kgg}}\PE\left[\left(1+u_f^2\epsilon^2+\frac{1}{u_f^{2}\epsilon^{2}}\right)\tHS^2-\tHS^4\right]\,.
\end{aligned}
\end{equation}
Summing over the possible values of $\epsilon$ we get
\begin{equation}\label{eqn:HSsu2}
\HS_{\mathbb{Z}_{\kgg}}^{\mathfrak{su}(2)}(\tHS,u_f) = \begin{cases}\PE\left[\left(1+u_f^2+u_f^{-2}\right)\tHS^2-\tHS^4\right]&{\kgg}=1\\[10pt]
\PE\left[\tHS^2+\left(u_f^{\kgg}+u_f^{-{\kgg}}\right)\tHS^{\kgg}-\tHS^{2{\kgg}}\right]&{\kgg}=2,4,6\\[10pt]
\PE\left[\tHS^2+\left(u_f^6+u_f^{-6}\right)\tHS^6-\tHS^{12}\right]&{\kgg}=3\\ 
\end{cases}\, .
\end{equation}
We again define the generators 
\begin{equation}
W^+=\frac{1}{2}\tr Y^{\kgg}\,,\quad W^-=\frac{1}{2}\tr Z^{\kgg}\,,\quad J=\frac{1}{2}\tr YZ\,.
\end{equation}
For ${\kgg}\in\{2,4,6\}$ we have $W^+W^-=J^{\kgg}$ and the topology of the Higgs branch is $\mathbb{C}^2/\mathbb{Z}_{\kgg}$. The ${\kgg}=1$ case is the same as ${\kgg}=2$. The ${\kgg}=3$ case is also the same as ${\kgg}=6$. 
\begin{comment}
Moreover, it is important to note that the $\kgg=3$ case cannot be obtained from the usual $\mathcal{N}=4$ theories since, when one considers non-local operators, the S-duality group is reduced to a subgroup of $SL(2,\mathbb{Z})$ , however it is only physical for discrete gauging of the `new' $\mathcal{N}=4$ theory \cite{Argyres:2016yzz} and the exact interpretation therefore of the ${\kgg}=3$ discrete gauging is still somewhat unclear. 
\end{comment}
In terms of $\mathfrak{su}(2,2|3)$ multiplets, after discarding the ${\kgg}=3$ case, we again have
\begin{equation}
\hat{\mathcal{B}}_{[{\kgg},0]}\hat{\mathcal{B}}_{[0,{\kgg}]}\color{black}\sim\left(\hat{\mathcal{B}}_{[1,1]}\color{black}\right)^{\kgg}\,.
\end{equation}
We again find agreement with \cite{Nishinaka:2016hbw,Lemos:2016xke}.

\section{Higher rank theories}
\label{sec:rankN}
\renewcommand{\arraystretch}{1.2}
Having studied in detail the rank one theories we now turn our attention to $\mathbb{Z}_n$ discrete gauging of higher rank theories. We limit most of our attention to the cases of $\mathfrak{g}=AD$ and $\mathfrak{g}=\mathfrak{u}(N)$ where the S-duality group \eqref{eqn:Sdual} acting on local operators is given by $SL(2,\mathbb{Z})$. In general the computation of the full discretely gauged index (\ref{eqn:discgaugepro}) for  $\mathfrak{g}=\mathfrak{u}(N),A,D$ is very difficult to perform. Therefore, also for this class of theories, we decide to focus our attention only on the Coulomb branch limit of the index (\ref{eqn:SCICB}) and on the Higgs branch Hilbert series (\ref{eq:HIggsHS}). For the Hilbert series we only explicitly present the rank $2$ cases. In the final subsection we will discuss the Coulomb branch index for the cases $\mathfrak{g}=E_6,E_7,E_8$. 
\subsection{$\mathfrak{g}=\mathfrak{u}(N)$} 
\subsubsection*{Coulomb branch limit}
Let us study the Coulomb branch limit \eqref{eqn:CBlimit}. Applying \eqref{eqn:SCICBdeg} we find
\begin{equation}\label{eqn:uNCBIndex}
\mathcal{I}_{\mathbb{Z}_{\kgg},\text{CB}}^{\mathfrak{u}(N)}(x)=\frac{1}{|\mathbb{Z}_{\kgg}|}\sum_{\epsilon\in\mathbb{Z}_{\kgg}}\PE\left[\sum_{j=1}^N\epsilon^jx^j\right]\,.
\end{equation}
We list a few cases for low rank. We define for ${\kgg}=1$ the generators of $CB_{\mathfrak{u}(N)}$ to be $u_j=\frac{1}{j}\tr X^j$. For $N=2$ we collate the results for the Coulomb branch index below 
\begin{center}
\begin{tabular}{|c|c|c|c|c|c|}
\hline
$\mathcal{I}_{\mathbb{Z}_{\kgg},\text{CB}}^{\mathfrak{u}(2)}(x)$&${\kgg}$&Generators&Relation&Topology\\[2pt] \hline
$\PE\left[x+x^2\right]$ & $1$ &$u_1,u_2$& $\diagup$ &$\mathbb{C}^2$\\[2pt] \hline
$\PE\left[2x^2\right]$ & $2$ &$\widetilde{u}_1=u_1^2$, $u_2$& $\diagup$ &$\mathbb{C}^2$\\\hline
$\PE\left[2x^3+x^6-x^9\right]$&$3$&$\widetilde{u}_1=u_1^3$, $\widetilde{u}_2=u_1u_2$, $\widetilde{u}_3=u_2^3$&$\widetilde{u}_1\widetilde{u}_3=\widetilde{u}_2^3$&$\mathbb{C}^2/\mathbb{Z}_3$\\\hline
$\PE\left[3x^4-x^8\right]$&$4$&$\widetilde{u}_1=u_1^4$, $\widetilde{u}_2=u_2^2$, $\widetilde{u}_3=u_1^2u_2$&$\widetilde{u}_1\widetilde{u}_2=\widetilde{u}_3^2$&$\mathbb{C}^2/\mathbb{Z}_2$\\\hline
$(1+2x^6)\PE\left[2x^6\right]$&$6$&\multicolumn{3}{c|}{Not complete intersection}\\\hline
\end{tabular}
\end{center}
By $\diagup$ we mean that the corresponding variety is freely generated with no relation. For ${\kgg}=3,4$ $CB_{\mathfrak{u}(2),\kgg}$ is not freely generated. Moreover for ${\kgg}=6$ we find that Coulomb branch is not a complete intersection. This is in agreement with the expectation that we outlined above \eqref{eqn:CBquotient}. The dimension of Coulomb branch, as a complex manifold, is given by applying \eqref{eqn:CBdim} and $\dim_{\mathbb{C}} CB_{\mathfrak{u}(2),\kgg}=2$ in each case.
\begin{comment}
Since we expect that the Plethystic logarithm of the Coulomb branch index is counting the number of generators of the Coulomb branch chiral ring, i.e. the number of independent $\mathcal{E}_{r,(0,0)}$ multiplets, minus relations. Let us examine this case in a bit more detail. 
\end{comment} 
For the case when $CB_{\mathfrak{u}(N),\kgg}$ is non-planar but a complete intersection one can easily read off the generators and relation. Conversely when it is not a complete intersection some more effort is required. The expansion of the Plethystic logarithm of the $\kgg=6$ Coulomb branch index reads
\begin{equation}\label{eqn:u2CBPlog}
\PLog\left[\mathcal{I}^{\mathfrak{u}(2)}_{\mathbb{Z}_6,\text{CB}}(x)\right]=4 x^6 - 3 x^{12} + 2 x^{18} +\mathcal{O}(x^{24})\,.
\end{equation} 
The generators at $x^6$ are 
\begin{equation}\label{eqn:u2CBgenerators}
\widetilde{u}_1=u_2^3\,,\quad\widetilde{u}_2=u_1^6\,,\quad\widetilde{u}_3=u_2u_1^4\,,\quad\widetilde{u}_4=u_2^2u_1^2\,,
\end{equation} 
they are primaries of the multiplets $\mathcal{E}_{6,(0,0)}$. 
\begin{comment}
Most of the higher order terms in the expansion of the Plethystic logarithm are redundant. Let us rewrite
\begin{equation}\label{eqn:u2CBrewrite}
\mathcal{I}_{\mathbb{Z}_6,\text{CB}}^{\mathfrak{u}(2)}(x)=\frac{1-3x^{12}+2x^{18}}{\left(1-x^6\right)^4}\,.
\end{equation} 
The four generators are encoded in the denominator of \eqref{eqn:u2CBrewrite}. The numerator \eqref{eqn:u2CBrewrite} should therefore encode relations and Hilbert syzygies.
\end{comment}
There are three relations at $x^{12}$ 
\begin{equation}\label{eqn:u2CBrelations}
I_1:\widetilde{u}_1\widetilde{u}_2-\widetilde{u}_3\widetilde{u}_4=0\,,\quad I_2:\widetilde{u}_4^2-\widetilde{u}_3\widetilde{u}_1=0\,,\quad I_3:\widetilde{u}_3^2-\widetilde{u}_2\widetilde{u}_4=0\,.
\end{equation}
However these relations are not all independent; at $x^{18}$ we have syzygies
\begin{equation}\label{eqn:u2CBsyzygies}
\widetilde{u}_3I_1+\widetilde{u}_2I_2+\widetilde{u}_4 I_3\equiv0\,,\quad \widetilde{u}_4 I_1+\widetilde{u}_3 I_2+\widetilde{u}_1 I_3\equiv0\,.
\end{equation}
Generally the moduli space should be characterised by \eqref{eqn:u2CBgenerators}, \eqref{eqn:u2CBrelations} and \eqref{eqn:u2CBsyzygies}.
For $N=3$ the Coulomb branch index is given by
\vspace{0.1cm}
\begin{center}
\begin{tabular}{|c|c|c|c|c|c|}
\hline
$\mathcal{I}_{\mathbb{Z}_{\kgg},\text{CB}}^{\mathfrak{u}(3)}(x)$&${\kgg}$&Generators&Relation&Topology\\\hline
$\PE\left[x+x^2+x^3\right]$ & $1$ &$u_1,u_2,u_3$&$\diagup$&$\mathbb{C}^3$\\\hline
\multirow{ 2}{*}{$\PE\left[2x^2+x^4+x^6-x^8\right]$} & \multirow{ 2}{*}{$2$} & $\widetilde{u}_1=u_1^2$, $u_2$, &\multirow{ 2}{*}{$\widetilde{u}_1\widetilde{u}_3=\widetilde{u}_2^2$}&\multirow{ 2}{*}{$\mathbb{C}\times\mathbb{C}^2/\mathbb{Z}_2$}\\
&&$\widetilde{u}_2=u_1u_3$, $\widetilde{u}_3=u_3^2$&&\\\hline
\multirow{ 2}{*}{$\PE\left[3x^3+x^6-x^9\right]$}&\multirow{ 2}{*}{$3$}&$\widetilde{u}_1=u_1^3$, $\widetilde{u}_2=u_1u_2$,&\multirow{ 2}{*}{$\widetilde{u}_1\widetilde{u}_3=\widetilde{u}_2^3$}&\multirow{ 2}{*}{$\mathbb{C}\times\mathbb{C}^2/\mathbb{Z}_3$}\\
&& $u_3$, $\widetilde{u}_3=u_2^3$&&\\\hline
$\frac{(1+x^4)(1+x^4+2x^8)}{\left(1-x^4\right)^3(1+x^4+x^8)}$&$4$&\multicolumn{3}{c|}{Not complete intersection}\\\hline
$(1+4x^6+x^{12})\PE\left[3x^6\right]$&$6$&\multicolumn{3}{c|}{Not complete intersection}\\\hline
\end{tabular}
\end{center}
\vspace{0.1cm}
For $N=4$ the Coulomb branch index is given by
\vspace{0.1cm}
\begin{center}
\begin{tabular}{|c|c|c|c|c|c|}
\hline
$\mathcal{I}_{\mathbb{Z}_{\kgg},\text{CB}}^{\mathfrak{u}(4)}(x)$&${\kgg}$&Generators&Relation&Topology\\\hline
$\PE\left[x+x^2+x^3+x^4\right]$ & $1$ &$u_1,u_2,u_3,u_4$&$\diagup$&$\mathbb{C}^4$\\\hline
\multirow{ 2}{*}{$\PE\left[2x^2+2x^4+x^6-x^8\right]$} & \multirow{ 2}{*}{$2$}&$u_2$, $\widetilde{u}_1=u_1^2$, $u_4$, &\multirow{ 2}{*}{$\widetilde{u}_1\widetilde{u}_3=\widetilde{u}_2^2$}&\multirow{ 2}{*}{$\mathbb{C}^2\times\mathbb{C}^2/\mathbb{Z}_2$}\\
&&$\widetilde{u}_2=u_1u_3$, $\widetilde{u}_3=u_3^2$&&\\\hline
$\frac{(1+x^3+x^6)(1+2x^6)}{(1-x^3)^4(1+x^3)^2(1+x^6)}$&$3$&\multicolumn{3}{c|}{Not complete intersection}\\\hline
$\frac{(1+x^4)(1+x^4+2x^8)}{\left(1-x^4\right)^4(1+x^4+x^8)}$&$4$&\multicolumn{3}{c|}{Not complete intersection}\\\hline
$\frac{(1+2x^6)(1+4x^6+x^{12})}{\left(1-x^6\right)^4(1+x^6)}$&$6$&\multicolumn{3}{c|}{Not complete intersection}\\\hline
\end{tabular}
\end{center}
\vspace{0.1cm}
We would like to point out that the dimension formula \eqref{eqn:CBdim} is in perfect agreement with the above results.
We checked up to $N=60$ and order $x^{70}$ that $CB_{\mathfrak{u}(N),n}$ for $\kgg\geq2$ is not a complete intersection for all $N\geq 5$. In principle the analysis that we performed \eqref{eqn:u2CBPlog} - \eqref{eqn:u2CBsyzygies} can be repeated for each case, however doing so is beyond the current scope of this article. Further note that for each $N$ and ${\kgg}\geq3$ we do not have Coulomb branch operators of dimension one or two, implying that we indeed have genuine $\mathcal{N}=3$ supersymmetry \cite{Aharony:2015oyb}. 

\subsubsection*{Higgs branch Hilbert series}
Let us now analyse the Higgs branch for these theories. %Beyond $N\geq3$ the computations become rather unmanageable since letter counting cannot be used.
We restrict our attention to the case $\mathfrak{g}=\mathfrak{u}(2)$. Using \textit{Macaulay2} and performing the integration over $U(2)$ gauge group the Higgs branch Hilbert series reads
\begin{equation}\label{eqn:HSU2}
\HS_{\mathbb{Z}_{\kgg}}^{\mathfrak{u}(2)}(\tHS,u_f) = \frac{1}{|\mathbb{Z}_{\kgg}|}\sum_{\epsilon \in \mathbb{Z}_{\kgg}}\PE\left[\left(\epsilon u_f+\epsilon^{-1} u_f^{-1}\right)\tHS+\left(1+\epsilon^2 u_f^2+\epsilon^{-2} u_f^{-2}\right)\tHS^2-\tHS^4 \right] \, .
\end{equation}
After performing the sum over $\mathbb{Z}_{\kgg}$ \eqref{eqn:HSU2} becomes
\begin{equation}
\HS_{\mathbb{Z}_{\kgg}}^{\mathfrak{u}(2)}(\tHS,u_f)=\begin{cases}\PE\left[(u_f+u_f^{-1})\tHS+(1+u_f^2+u_f^{-2})\tHS^2-\tHS^4\right]&{\kgg}=1\\[12pt]
\PE\left[2(1+u_f^2+u_f^{-2})\tHS^2-2\tHS^4\right]&{\kgg}=2\\[12pt]
\frac{(1 + \tHS^2) \left(\tHS^6 (u_f^6 + u_f^{-6}) + 
   \tHS^3 (1 + \tHS^2) (1 + \tHS^2 + \tHS^4) (u_f^3 + u_f^{-3}) + 1 + \tHS^2 + 4 \tHS^4 + 
     \tHS^6 + 4 \tHS^8 + \tHS^{10} + \tHS^{12} \right)}{\left(1 + \tHS^6 - 
   \tHS^3 (u_f^3 + u_f^{-3})\right)^2 \left(1 + \tHS^6 + \tHS^3 (u_f^3 + u_f^{-3})\right)}&{\kgg}=3\\[12pt]
   \frac{(1 + \tHS^2)^2 (u_f^4 + \tHS^4 (1 + (4 + \tHS^4) u_f^4 + u_f^8))}{(1 + 
  \tHS^8  - \tHS^4 ( u_f^4 + u_f^{-4}))}^2&{\kgg}=4\\[12pt]
 \frac{(1 + \tHS^2)^2  (2 \tHS^6 (1 + \tHS^4) + (1 + 4 \tHS^4 + 9 \tHS^8 + 4 \tHS^{12} + 
      \tHS^{16}) u_f^6 + 2 \tHS^6 (1 + \tHS^4) u_f^{12})}{(1 + \tHS^{12}  - 
  \tHS^6 (u_f^6 + u_f^{-6}))^2} &{\kgg}=6
\end{cases}
\end{equation}
When ${\kgg}=1,2$ we get a complete intersection. Moreover, in an expansion around $\tHS$ the dependence on $u_f$ in \eqref{eqn:HSU2} arranges itself into characters of $SU(2)$ implying that the $U(1)_f$ isometry of the Higgs branch is enhanced to $SU(2)_f$ for these theories. This is of course due to the fact that supersymmetry is enhanced to $\mathcal{N}=4$ for ${\kgg}=1,2$. For ${\kgg}=3,4,6$ we do not have complete intersections, nonetheless we may identify the first generators and their relation. Moreover, for each ${\kgg}$, by applying the dimension formula \eqref{eqn:dimHB} we find that the Higgs branch is a manifold of complex dimension four.
We define 
\begin{equation}
W_{j,m}=\frac{1}{j+m}\tr Y^jZ^m\,.
\end{equation}
For ${\kgg}=1$, by taking the Plethystic logarithm of \eqref{eqn:HSU2}, we find that the Higgs branch is generated by the $W_{j,0}$, $W_{0,j}$ for $j=1,2$ and $W_{1,1}$. There is a relation of dimension $4$ between them given by $2W_{1,1}(2W_{1,1}-W_{0,1}W_{1,0})+W_{0,1}^2W_{2,0}+W_{1,0}^2W_{0,2}=0$. The topology is $HB_{\mathfrak{u}(2),1}\iso\Sym^2\left(\mathbb{C}^2\right)$ \cite{Nakajima:2003pg,Feng:2007ur}.
%\begin{equation}
%4J^2-4W_2V_2+4W_1V_1J+W_2{V_1}^2+V_2{W_1}^2=0\,.
%\end{equation}
For ${\kgg}=2$ the Higgs branch is generated by $W_{0,2}$, $W_{2,0}$ $W_{1,1}$ , $\widetilde{W}={W_{1,0}}^2$, $\widetilde{V}={W_{0,1}}^2$ and $\widetilde{J}=W_{1,0}W_{0,1}$ and there are two relations of dimension $4$. The topology is $HB_{\mathfrak{u}(2),2}\iso\mathbb{C}^2/\mathbb{Z}_2\times\mathbb{C}^2/\mathbb{Z}_2$.
%\begin{equation}
%4J^2-4W_2V_2-J\widetilde{J}+\widetilde{W}_2\widetilde{V}_2+2W_2\widetilde{V}_2+V_2\widetilde{W}_2=0\,,\quad \widetilde{W}_2\widetilde{V}_2=\widetilde{J}^2\,.
%\end{equation}
At ${\kgg}=3$ we do not get a complete intersection, nevertheless we can expand the Plethystic logarithm of \eqref{eqn:HSU2}
\begin{equation}\label{eqn:HSU2Z3Plog}
\begin{aligned}
\PLog\left[\HS_{\mathbb{Z}_3}^{\mathfrak{u}(2)}(\tHS,u_f)\right]=&2 \tHS^2+2 \tHS^3 \left(u_f^3+u_f^{-3}\right)+2 \tHS^4+\tHS^5 \left(u_f^3+u_f^{-3}\right)\\
&+\tHS^6
   \left(u_f^6+u_f^{-6}-4\right)+\mathcal{O}\left(\tHS^7\right)\,.
  \end{aligned}
\end{equation}
The generators are $W_{1,1}$, $\widetilde{J}_1=W_{0,1}W_{1,0}$, $\widetilde{W}_1={W_{1,0}}^3$, $\widetilde{V}_1={W_{0,1}}^3$, $\widetilde{W}_2=W_{2,0}W_{1,0}$, $\widetilde{V}_2=W_{0,2}W_{0,1}$, $\widetilde{J}_2=W_{2,0}W_{0,2}$, $W_{2,2}$, $\widetilde{W}_3=W_{2,0}^2W_{0,1}$, $\widetilde{V}_3={W_{0,2}}^2W_{1,0}$, $\widetilde{W}_4={W_{0,2}}^3$ and $\widetilde{V}_4={W_{2,0}}^3$. There are four relations of dimension six between them. In terms of $\mathfrak{su}(2,2|3)$ mutiplets these have the correct quantum numbers to be
\begin{equation}
\begin{aligned}\label{eqn:u2z3HB1}
&\hat{\mathcal{B}}_{[1,1]}\,,\quad \hat{\mathcal{B}}_{[1,1]}\,,\quad \hat{\mathcal{B}}_{[3,0]} \,,\quad {\hat{\mathcal{B}}_{[0,3]}}\,,\quad \hat{\mathcal{B}}_{[3,0]} \,,\quad {\hat{\mathcal{B}}_{[0,3]}}\,,\\ &{\hat{\mathcal{B}}_{[2,2]}}\,,\quad {\hat{\mathcal{B}}_{[2,2]}}\,,\quad {\hat{\mathcal{B}}_{[4,1]}}\,,\quad {\hat{\mathcal{B}}_{[1,4]}}\,,\quad {\hat{\mathcal{B}}_{[6,0]}}\,,\quad {\hat{\mathcal{B}}_{[0,6]}}\,.
\end{aligned}
\end{equation} 
%Higher order terms in the expansion of Plethystic logarithm represent relations of higher order and syzygies.
Note that, using \eqref{eqn:Bnmdecomp}-\eqref{eqn:B0n}, it is easily checked that \eqref{eqn:u2z3HB1} agrees with the spectrum of Coulomb branch operators that we found for the $\mathfrak{u}(2)$ $\mathbb{Z}_{\kgg=3}$ theory \eqref{eqn:uNCBIndex}. Note that in \eqref{eqn:HSU2Z3Plog} two generators appear which have the correct quantum numbers to belong to $\hat{\mathcal{B}}_{[1,1]}$ multiplets. This implies that the theory contains two conserved spin two currents (which lie inside $\hat{\mathcal{C}}_{0,(0,0)}$ multiplets in $\mathcal{N}=2$ language).

At ${\kgg}=4,6$ we again do not get complete intersection varieties. One can perform a similar analysis for those cases as we did for ${\kgg}=3$.
%At $k=4$ we again do not get a complete intersection. 
%\begin{equation}
%\begin{split}
%\textrm{PLog}[\textrm{HS}_{\mathbb{Z}_4}^{\mathfrak{u}(2)}(t,x)] & =2 t^2+t^4 \left(3 x^4+\frac{3}{x^4}+2\right)+t^8 \left(-x^8-\frac{1}{x^8}-4 x^4-\frac{4}{x^4}-10\right)+\\
%& t^{12} \left(4
%   x^8+\frac{4}{x^8}+16 x^4+\frac{16}{x^4}+24\right)+O\left(t^{13}\right)
%\end{split}
%\end{equation}
%\paragraph{$k=6$}
%We do not get a complete intersection
%\begin{equation}
%\begin{split}
%\textrm{PLog}[\textrm{HS}(t,x)_{\mathbb{Z}_6}^{\mathfrak{u}(2)}] & = 2 t^2+2 %t^4+t^6 \left(4 x^6+\frac{4}{x^6}\right)-t^8+t^{10} \left(-6 x^6-\frac{6}{x^6}\right)+\\
%& t^{12} \left(-3
%   x^{12}-\frac{3}{x^{12}}-16\right)+O\left(t^{13}\right)
%   \end{split}
%\end{equation}

\subsection{$\mathfrak{g}=\mathfrak{su}(N+1)$} 
\subsubsection*{Coulomb branch limit}
Let us study the Coulomb branch limit. From \eqref{eqn:SCICBdeg} we have
\begin{equation}\label{eqn:CBsuN}
\mathcal{I}_{\mathbb{Z}_{\kgg},\text{CB}}^{\mathfrak{su}(N+1)}(x)=\frac{1}{|\mathbb{Z}_{\kgg}|}\sum_{\epsilon\in\mathbb{Z}_{\kgg}}\PE\left[\sum_{j=2}^{N+1}\epsilon^jx^j\right]\,.
\end{equation}
Let us examine a few cases for low rank. We define the generators of $CB_{\mathfrak{su}(N+1)}$ for the parent theory to be given by $u_j=\frac{1}{j}\tr X^j$. For $N+1=3$ we have
\vspace{0.1cm}
\begin{center}
\begin{tabular}{|c|c|c|c|c|c|}
\hline
$\mathcal{I}_{\mathbb{Z}_{\kgg},\text{CB}}^{\mathfrak{su}(3)}(x)$&${\kgg}$&Generators&Relation&Topology\\\hline
$\PE\left[x^2+x^3\right]$ & $1$&$u_2$, $u_3$&$\diagup$&$\mathbb{C}^2$\\\hline
$\PE\left[x^2+x^6\right]$ & $2$ & $u_2$, $\widetilde{u}_1=u_3^2$&$\diagup$&$\mathbb{C}^2$\\\hline
$\PE\left[x^3+x^6\right]$ & $3$& $u_3$, $\widetilde{u}_1=u_2^3$&$\diagup$&$\mathbb{C}^2$\\\hline
$\PE\left[x^4+x^8+x^{12}-x^{16}\right]$ & $4$ & $\widetilde{u}_1=u_2^2$, $\widetilde{u}_2=u_2u_3^2$, $\widetilde{u}_3=u_3^4$&$\widetilde{u}_1\widetilde{u}_3=\widetilde{u}_2^2$&$\mathbb{C}^2/\mathbb{Z}_2$\\\hline
$\PE\left[2x^6\right]$ & $6$&$\widetilde{u}_1=u_2^3$, $\widetilde{u}_2=u_3^2$&$\diagup$&$\mathbb{C}^2$\\\hline
\end{tabular}
\end{center}
\vspace{0.1cm}
When $\kgg=1,2,3,6$ $CB_{\mathfrak{su}(3),\kgg}$ is freely generated, in agreement with our discussion above \eqref{eqn:CBquotient}. For $N+1=4$ we have
\vspace{0.1cm}
\begin{center}
\begin{tabular}{|c|c|c|c|c|c|}
\hline
$\mathcal{I}_{\mathbb{Z}_{\kgg},\text{CB}}^{\mathfrak{su}(4)}(x)$&${\kgg}$&Generators&Relation&Topology\\\hline
$\PE\left[x^2+x^3+x^4\right]$ & $1$ &$u_2$, $u_3$, $u_4$&$\diagup$&$\mathbb{C}^3$\\\hline
$\PE\left[x^2+x^4+x^6\right]$ & $2$ & $u_2$, $u_4$, $\widetilde{u}_1=u_2^3$&$\diagup$&$\mathbb{C}^3$\\\hline
\multirow{ 2}{*}{$\PE\left[x^3+2x^6+x^{12}-x^{18}\right]$} & \multirow{ 2}{*}{$3$} & $u_3$, $\widetilde{u}_1=u_2^3$,&\multirow{ 2}{*}{$\widetilde{u}_1\widetilde{u}_3=\widetilde{u}_2^3$}&\multirow{ 2}{*}{$\mathbb{C}\times\mathbb{C}^2/\mathbb{Z}_3$}\\
&&$\widetilde{u}_2=u_2u_4$, $\widetilde{u}_3=u_4^3$&&\\\hline
\multirow{ 2}{*}{$\PE\left[2x^4+x^8+x^{12}-x^{16}\right]$} & \multirow{ 2}{*}{$4$} & $\widetilde{u}_1=u_2^2$, $u_4$,&\multirow{ 2}{*}{$\widetilde{u}_1\widetilde{u}_3=\widetilde{u}_2^2$}&\multirow{ 2}{*}{$\mathbb{C}\times\mathbb{C}^2/\mathbb{Z}_2$}\\
&&$\widetilde{u}_2=u_2u_3^2$, $\widetilde{u}_3=u_3^4$&&\\\hline
\multirow{ 2}{*}{$\PE\left[3x^6+x^{12}-x^{18}\right]$} & \multirow{ 2}{*}{$6$} & $\widetilde{u}_1=u_2^3$, $\widetilde{u}_2=u_3^2$,&\multirow{ 2}{*}{$\widetilde{u}_1\widetilde{u}_4=\widetilde{u}_3^3$}&\multirow{ 2}{*}{$\mathbb{C}\times\mathbb{C}^2/\mathbb{Z}_3$}\\
&&$\widetilde{u}_3=u_2u_4$, $\widetilde{u}_4=u_4^3$&&\\\hline
\end{tabular}
\end{center}
\vspace{0.1cm}
For $N+1=5$ we have
\vspace{0.1cm}
\begin{center}
\begin{tabular}{|c|c|c|c|c|c|}
\hline
$\mathcal{I}_{\mathbb{Z}_{\kgg},\text{CB}}^{\mathfrak{su}(5)}(x)$&${\kgg}$&Generators&Relation&Topology\\\hline
$\PE\left[\sum_{A=2}^5x^A\right]$ & $1$ &$u_2,u_3,u_4,u_5$&$\diagup$&$\mathbb{C}^4$\\\hline
\multirow{ 2}{*}{$\PE\left[\sum_{A=1}^5x^{2A}-x^{16}\right]$} & \multirow{ 2}{*}{$2$} & $u_2$, $u_4$, $\widetilde{u}_1=u_3^2$,&\multirow{ 2}{*}{$\widetilde{u}_3\widetilde{u}_1=\widetilde{u}_2^2$}&\multirow{ 2}{*}{$\mathbb{C}^2\times\mathbb{C}^2/\mathbb{Z}_2$}\\
&&$\widetilde{u}_2=u_3u_5$, $\widetilde{u}_3=u_5^2$&&\\\hline
$\frac{1+x^6+2x^9+2x^{12}+x^{15}+2x^{18}}{(1-x^3)^4(1+x^3)^2(1+(x^3+x^6+x^9)(1+x^3+x^9))}$ & $3$&\multicolumn{3}{c|}{Not complete intersection}\\\hline
$\frac{(1+x^8)(1+x^8+x^{12}+x^{16})}{(1-x^4)^4(1+(x^4+x^8)(2+2x^4+2x^8+x^{12}+x^{16}))}$ & $4$&\multicolumn{3}{c|}{Not complete intersection}\\\hline
$\frac{1+x^6+4x^{12}+4x^{18}+3x^{24}+3x^{30}+2x^{36}}{(1-x^6)^4(1+2x^6+2x^{12}+2x^{18}+2x^{24}+2x^{30})}$ & $6$&\multicolumn{3}{c|}{Not complete intersection}\\\hline
\end{tabular}
\end{center}
Out of the theories with $N+1>5$ we find that, apart from ${\kgg}=2$, $N+1=6$, the Coulomb branch for ${\kgg}=2,3,4,6$ is never a complete intersection. We checked this up to $N=60$ and $x^{70}$. In each case the dimension formula \eqref{eqn:CBdim} holds and is equal to $N$ as expected. In the cases where the moduli space is not a complete intersection variety the analysis that we demonstrated \eqref{eqn:u2CBPlog} - \eqref{eqn:u2CBsyzygies} can, in principle, be repeated. Again, for each $N$, with ${\kgg}\geq3$ we do not have Coulomb branch operators of dimension one or two implying genuine $\mathcal{N}=3$ supersymmetry \cite{Aharony:2015oyb}. 

\subsubsection*{Higgs branch Hilbert series}
Let us turn to analysing the Higgs branch for these theories. We restrict ourselves only to the case $\mathfrak{g}=\mathfrak{su}(3)$. Using \textit{Macaluay2} and performing the integration over the gauge group the Higgs branch Hilbert series reads
\begin{equation}
\label{eq:hsu3}
\begin{aligned}
&\HS_{\mathbb{Z}_{\kgg}}^{\mathfrak{su}(3)}(\tHS,u_f)= \frac{1}{{\kgg}}\sum_{\epsilon \in \mathbb{Z}_{\kgg}}\frac{1+\tHS^2+ \left(u_f \epsilon +\frac{1}{u_f \epsilon }\right)\tHS^3+\tHS^4+\tHS^6}{\left(1-\frac{\tHS^2}{u_f^2 \epsilon ^2}\right)\left(1-\tHS^2u_f^2\epsilon^2\right)\left(1-\frac{\tHS^3}{u_f^3\epsilon^3}\right)\left(1-\tHS^3u_f^3\epsilon^3\right) }\,.
\end{aligned}
\end{equation}
We find that, for all ${\kgg}$, the corresponding moduli space is never a complete intersection. Moreover, applying \eqref{eqn:dimHB}, we find that in each case the Higgs branch is of complex dimension four.
A complete analysis of the Higgs branches of these theories is beyond the scope of this paper. However, as we did for the $\mathfrak{u}(2)$ case we would like to demonstrate with an example. The generators of the parent $({\kgg}=1)$ theory are $W_{j,0}$ and $W_{0,j}$ for $j\in\{2,3\}$, $W_{1,1}$, $W_{2,1}$ and $W_{1,2}$ where, as before,
\begin{equation}
W_{j,m}=\frac{1}{j+m}\tr Y^jZ^m\,.
\end{equation}
As an example let us expand the Plethystic logarithm of \eqref{eq:hsu3} for ${\kgg}=3$
\begin{equation}
\begin{aligned}
\PLog\left[\HS_{\mathbb{Z}_3}^{\mathfrak{su}(3)}(\tHS,u_f)\right]= &\tHS^2 + (u_f^{-3} + u_f^3) \tHS^3 + \tHS^4 + (u_f^{-3} + u_f^3) \tHS^5 + (u_f^{-6} + 
    u_f^6) \tHS^6 - \tHS^8 \\&- (u_f^{-6} + u_f^6) \tHS^9
    -(1+u_f^{-6}+u_f^6)\tHS^{10}+\mathcal{O}(\tHS^{11})\,.
\end{aligned}
\end{equation} 
The generators are $W_{1,1}$, $W_{3,0}$, $W_{0,3}$, $\widetilde{J}=W_{2,0}W_{0,2}$, $\widetilde{W}_1=W_{2,1}W_{2,0}$, $\widetilde{V}_1=W_{1,2}W_{0,2}$, $\widetilde{W}_2={W_{2,0}}^3$ and $\widetilde{V}_2={W_{0,2}}^3$. In terms of $\mathfrak{su}(2,2|3)$ multiplets these have the correct quantum numbers to correspond to
\begin{equation}
{\hat{\mathcal{B}}_{[1,1]}}\,,\quad {\hat{\mathcal{B}}_{[3,0]}}\,,\quad {\hat{\mathcal{B}}_{[0,3]}}\,,\quad {\hat{\mathcal{B}}_{[2,2]}}\,,\quad {\hat{\mathcal{B}}_{[4,1]}}\,,\quad {\hat{\mathcal{B}}_{[1,4]}}\,,\quad {\hat{\mathcal{B}}_{[6,0]}}\,,\quad {\hat{\mathcal{B}}_{[0,6]}}\,.
\end{equation}

\subsection{$\mathfrak{g}=\mathfrak{so}(2N)$} 
\subsubsection*{Coulomb branch limit}
The Coulomb branch limit \eqref{eqn:SCICBdeg} reads
\begin{equation}
\mathcal{I}_{\mathbb{Z}_{\kgg},\text{CB}}^{\mathfrak{so}(2N)}(x)=\frac{1}{|\mathbb{Z}_{\kgg}|}\sum_{\epsilon\in\mathbb{Z}_{\kgg}}\PE\left[\epsilon^Nx^N+\sum_{j=1}^{N-1}\epsilon^{2j}x^{2j}\right]\,.
\end{equation}
We would like to discuss firstly the ${\kgg}=2$ case where there are two distinct cases. Namely when $N=2M$ or $N=2M-1$ for $M\in\mathbb{Z}$. 
Let the us choose a basis for the Coulomb branch chiral ring given by 
\begin{equation}
u_{2j}=\tr X^{2j}\,,\quad 1\leq j\leq N-1\quad\text{and}\quad \hat{u}_N=\Pf X \, ,
\end{equation}
where Pf denotes the Pfaffian. The dimensions of the above operator are $E(u_j)=2j$, $E(\hat{u}_N)=N$. When $\mathfrak{g}=\mathfrak{so}(4M)$ we can write $X=\diag\left(x_1\sigma_2,x_2\sigma_2,\dots,x_{2N}\sigma_2\right)$ then the $\mathbb{Z}_2$ acts by $r_2\newdot s_2:X\mapsto-X=g^{-1}Xg$ with $g=\diag\left(\sigma_3,\sigma_3,\dots,\sigma_3\right)\in SO(4M)$, where $\sigma_i$ denotes the Pauli matrices, and thus $r_2\cdot s_2$ is isomorphic to a gauge transformation and therefore the ${\kgg}=2$ case with $N=2M$ should lead to exactly the same theory as the ${\kgg}=1$ case. 
This is to be compared to the case when $\mathfrak{g}=\mathfrak{so}(4M-2)$. Writing $X=\diag\left(x_1\sigma_2,x_2\sigma_2,\dots,x_{2N-1}\sigma_2\right)$ as before we have $r_2\cdot s_2:X\mapsto-X=g^{-1}Xg$ now with $g=\diag\left(\sigma_3,\sigma_3,\dots,\sigma_3\right)\not\in SO(4N-2)$, infact, $g\in O(4N-2)$ and in this case the $\mathbb{Z}_2$ does generate a genuine global symmetry which, when gauged, will lead to a distinct theory. 
%\footnote{We should point out that the this does not violate the expectation that one should be able to obtain the $O(2N)$ theory as a discrete gauging of the $SO(2N)$ by it's $\mathbb{Z}_2$ central extension $SO(2N)\ltimes\mathbb{Z}_2$. Gauging the $\mathbb{Z}_2$ generated by $r_2\cdot s_2$ does not yield the $O(2N)$ theory (for $N$ odd), since it is not quite the correct $\mathbb{Z}_2$ one must gauge - the extention of $SO(2N)$ by $r_2\cdot s_2$ is a direct rather than semi-direct product. To obtain the $O(2N)$ theory via a discrete gauging of the $SO(2N)$ theory one must gauge the $\mathbb{Z}_2$ global symmetry generated by the element $\mathbb{Z}_2:X\mapsto h^{-1}Xh$ with $h=\diag\left(\sigma_3,\mathbb{I}_{2N-2}\right)\in O(2N)$ but $h\not\in SO(2N)$ and the extension of $SO(2N)$ by the $\mathbb{Z}_2$ generated by $h$ is indeed a semi-direct product. See e.g. \cite{Bourget:2017tmt,Bourget:2017sxr} for discussion on how this may be implemented at the level of the index/Hilbert series.}
Indeed we find that for $N=2M$
\begin{equation}
\mathcal{I}_{\mathbb{Z}_2,\text{CB}}^{\mathfrak{so}(4M)}(x)=\PE\left[x^{2M}+\sum_{j=1}^{2M-1}x^{2j}\right]=\mathcal{I}_{\mathbb{Z}_1,\text{CB}}^{\mathfrak{so}(4M)}(x)\,.
\end{equation}
On the other hand, for $N=2M-1$
\begin{equation}
\mathcal{I}_{\mathbb{Z}_2,\text{CB}}^{\mathfrak{so}(4M-2)}(x)=\PE\left[x^{4M-2}+\sum_{j=1}^{2M-2}x^{2j}\right]\,,
\end{equation}
and the new Coulomb branch operators are simply given by $u_2,u_4,\dots,u_{4M-4}$ and $\widetilde{u}=\left(\hat{u}_{2M-1}\right)^2=\det X$.
Let us now turn on the cases ${\kgg}=3,4,6$ for different values of $N$. In the following we collate the results that we found.\\
For $N=2$ we have
\begin{center}
\begin{tabular}{|c|c|c|c|c|}
\hline
$\mathcal{I}_{\mathbb{Z}_{\kgg},\text{CB}}^{\mathfrak{so}(4)}(x)$&${\kgg}$&Generators&Relation&Topology\\\hline
$(1+2x^6)\PE\left[1-2x^6\right]$ & $3$ &\multicolumn{3}{c|}{Not complete intersection}\\\hline
$\PE\left[3x^4-x^{8}\right]$ & $4$ & $\widetilde{u}_1=u_2\hat{u}_2$, $\widetilde{u}_2=u_2^2$, $\widetilde{u}_3=\hat{u}_2^2$&$\widetilde{u}_1^2=\widetilde{u}_2\widetilde{u}_3$&$\mathbb{C}^2/\mathbb{Z}_2$\\\hline
$(1+2x^6)\PE\left[1-2x^6\right]$ & $6$&\multicolumn{3}{c|}{Not complete intersection}\\\hline
\end{tabular}
\end{center}
Note that, since $\mathfrak{so}(4)\iso\mathfrak{su}(2)\oplus\mathfrak{su}(2)$, for $\kgg=1,2$ we have $\mathcal{I}_{\mathbb{Z}_{\kgg=1,2},\text{CB}}^{\mathfrak{so}(4)}=\left(\mathcal{I}_{\mathbb{Z}_{\kgg=1,2},\text{CB}}^{\mathfrak{su}(2)}\right)^2$. On the other hand, for $\kgg\geq3$, $\mathcal{I}_{\mathbb{Z}_{\kgg=3,4,6},\text{CB}}^{\mathfrak{so}(4)}\neq \left(\mathcal{I}_{\mathbb{Z}_{\kgg=3,4,6},\text{CB}}^{\mathfrak{su}(2)}\right)^2$. 
Since $\mathfrak{so}(6)\iso\mathfrak{su}(4)$ the Coulomb branch index for $N=3$ is the same as for the Coulomb branch index for the $\mathfrak{g}=\mathfrak{su}(4)$ theory \eqref{eqn:CBsuN} and therefore $\mathcal{I}_{\mathbb{Z}_{\kgg},\text{CB}}^{\mathfrak{so}(6)}(x)=\mathcal{I}_{\mathbb{Z}_{\kgg},\text{CB}}^{\mathfrak{su}(4)}(x)$.
\begin{comment}
\begin{center}
\begin{tabular}{|c|c|c|c|c|}
\hline
$\mathcal{I}_{\mathbb{Z}_{\kgg},\text{CB}}^{\mathfrak{so}(6)}(x)$&${\kgg}$&Generators&Relation&Topology\\\hline
\multirow{ 2}{*}{$\PE\left[x^3+2x^6+x^{12}-x^{18}\right]$} & \multirow{ 2}{*}{$3$} &$\hat{u}_3$, $\widetilde{u}_1=u_2^3$,
 &\multirow{ 2}{*}{$\widetilde{u}_1\widetilde{u}_3=\widetilde{u}_2^3$}&\multirow{ 2}{*}{$\mathbb{C}\times\mathbb{C}^2/\mathbb{Z}_3$}\\
 &&$\widetilde{u}_2=u_2u_4$, $\widetilde{u}_3=u_4^3$&&\\\hline
\multirow{ 2}{*}{$\PE\left[2x^4+x^{8}+x^{12}-x^{16}\right]$} & \multirow{ 2}{*}{$4$} & $u_4$, $\widetilde{u}_1=u_2^2$, &\multirow{ 2}{*}{$\widetilde{u}_2^2=\widetilde{u}_1\widetilde{u}_3$}&\multirow{ 2}{*}{$\mathbb{C}\times\mathbb{C}^2/\mathbb{Z}_2$}\\
&&$\widetilde{u}_2=\hat{u}_3^2u_2$, $\widetilde{u}_3=\hat{u}_3^4$&&\\\hline
\multirow{ 2}{*}{$\PE\left[3x^6+x^{12}-x^{18}\right]$ }&\multirow{ 2}{*}{$6$} &$\widetilde{u}_1=\hat{u}_3^2$, $\widetilde{u}_2=u_2u_4$,&\multirow{ 2}{*}{$\widetilde{u}^3_2=\widetilde{u}_3\widetilde{u}_4$}&\multirow{ 2}{*}{$\mathbb{C}\times\mathbb{C}^2/\mathbb{Z}_3$}\\
&& $\widetilde{u}_3=u_2^3$, $\widetilde{u}_4=u_4^3$&&\\\hline
\end{tabular}
\end{center}
\end{comment}
For $N=4$ we find
\begin{center}
\begin{tabular}{|c|c|c|c|c|c|}
\hline
$\mathcal{I}_{\mathbb{Z}_{\kgg},\text{CB}}^{\mathfrak{so}(8)}(x)$&${\kgg}$&Generators&Relation&Topology\\\hline
$\frac{1+2x^6+5x^{12}+x^{18}}{(1-x^6)^4(1+x^6)^2}$ & $3$ &\multicolumn{3}{c|}{Not complete intersection}\\\hline
\multirow{ 2}{*}{$\PE\left[3x^4+x^{8}+x^{12}-x^{16}\right]$} & \multirow{ 2}{*}{$4$}& $u_4$, $\hat{u}_4$, $\widetilde{u}_1=u_2^2$,&\multirow{ 2}{*}{$\widetilde{u}_2^2=\widetilde{u}_1\widetilde{u}_3$}&\multirow{ 2}{*}{$\mathbb{C}^2\times\mathbb{C}^2/\mathbb{Z}_2$}\\
&&$\widetilde{u}_2=u_2u_6$, $\widetilde{u}_3=u_6^2$&&\\\hline
$\frac{1+2x^6+5x^{12}+x^{18}}{(1-x^6)^4(1+x^6)^2}$ & $6$ &\multicolumn{3}{c|}{Not complete intersection}\\\hline
\end{tabular}
\end{center}
Out of the theories with $N>4$ we find that, apart from the ${\kgg}=2$ cases, which we discussed separately, the Coulomb branch for ${\kgg}=3,4,6$ is a not a complete intersection. We again checked this up to $N=60$ and $x^{70}$. In each case the dimension formula \eqref{eqn:CBdim} holds and the dimension is equal to $N$ as expected. With ${\kgg}\geq3$ we do not have Coulomb branch operators of dimension one or two, implying that we indeed have genuine $\mathcal{N}=3$ supersymmetry \cite{Aharony:2015oyb}. 

\subsubsection*{Higgs branch Hilbert series}
Using the software \textit{Macaulay2} we did the computation of the Higgs branch Hilbert series for the theory with Lie algebra $\mathfrak{g}=\mathfrak{so}(4) \iso \mathfrak{su}(2) \oplus \mathfrak{su}(2)$. After the integration over the gauge group we get
\begin{equation}
\HS_{\mathbb{Z}_{\kgg}}^{\mathfrak{so}(4)}(\tHS,u_f)=\frac{1}{|\mathbb{Z}_{\kgg}|}\sum_{\epsilon \in \mathbb{Z}_{\kgg}}\PE[2\tHS^2+2(\epsilon^2 u_f^2+\epsilon^{-2}u_f^{-2})\tHS^2-2\tHS^4] \, .
\end{equation}
We observe that the above Hilbert series has a pole of order four at $\tHS=1$ and therefore, by \eqref{eqn:dimHB}, the complex dimension of the Higgs branch is four. 
For ${\kgg}=1,2$ we get a complete intersection variety with Hilbert series
\begin{equation}
\HS_{\mathbb{Z}_1}^{\mathfrak{so}(4)}(\tHS,u_f) = \HS_{\mathbb{Z}_2}^{\mathfrak{so}(4)}(\tHS,u_f) = \textrm{PE}[2(1+u_f^2+u_f^{-2})\tHS^2 -2\tHS^4]= \left(\HS_{\mathbb{Z}_1}^{\mathfrak{su}(2)}(\tHS,u_f)\right)^2\, .
\end{equation}
At ${\kgg}=1,2$ it is clear that the Higgs branch moduli space is equal to two copies of the $\mathfrak{su}(2)$ case. We discussed that in Section \ref{sec:su2}. The topology of the moduli space is therefore $\mathbb{C}^2/\mathbb{Z}_2 \times \mathbb{C}^2/\mathbb{Z}_2$.
\begin{comment}
In order to characterize this moduli space let's introduce the fields
\begin{equation}\begin{split}
& Y_1 = \textrm{Diag}(y_1,-y_1,0,0), \ \ Y_2=\textrm{Diag}(0,0,y_2,-y_2),\\
& \ \ Z_1=\textrm{Diag}(z_1,-z_1,0,0), \ \ Z_2 =\textrm{Diag}(0,0,z_2,-z_2)
\end{split}\end{equation}
the generators at order $\tHS^2$ are
\begin{equation}
\textrm{Tr}[Y_1^2], \ \ \textrm{Tr}[Y_2^2], \ \ \textrm{Tr}[Z_1^2], \ \ \textrm{Tr}[Z_2^2],\ \ \textrm{Tr}[Y_1Z_1], \ \ \textrm{Tr}[Y_2Z_2] \, ,
\end{equation}
and satisfy the relations appearing at order $\tHS^4$
\begin{equation}
\textrm{Tr}[Z_1^2]\textrm{Tr}[Y_1^2]=\textrm{Tr}[Y_1Z_1], \ \ \textrm{Tr}[Z_2^2]\textrm{Tr}[Y_2^2]=\textrm{Tr}[Z_2Y_2] \, ,
\end{equation}
therefore the topology of this moduli space is $\mathbb{C}^2/\mathbb{Z}_2 \times \mathbb{C}^2/\mathbb{Z}_2$.
\end{comment}
For ${\kgg}=3,4,6$ we observe that the corresponding Hilbert series is not a complete intersection. Moreover the Hilbert series for ${\kgg}=3,6$ are equal.
\subsection{$\mathfrak{g}=E_N$} 
%The discretely gauged superconformal index for the exceptional gauge groups reads
%\begin{equation}
%\mathcal{I}^{E_N}_{\mathbb{Z}_{\kgg}}(x) = \frac{1}{|\mathbb{Z}_{\kgg}|}\sum_{\epsilon \in \mathbb{Z}_{\kgg}}\int d\mu_{\Exp E_N}(\mathbf{z}) \PE\left[i_{\kgg}(t,y,p,q,\epsilon)\chi^{\Exp E_N}_{\text{adj}}(\mathbf{z}) \right] \, .
%\end{equation}
In this subsection, since we can make use of \eqref{eqn:SCICBdeg}, we focus on the Coulomb branch limit of the index for $E_6,E_7$ and $ E_8$.
\paragraph{$\mathfrak{g}=E_6$}
The Coulomb branch index reads
\begin{equation}
\mathcal{I}_{\mathbb{Z}_{\kgg},\text{CB}}^{E_6}(x)=\frac{1}{|\mathbb{Z}_{\kgg}|}\sum_{\epsilon\in\mathbb{Z}_{\kgg}}\PE\left[\epsilon^2 x^2+\epsilon^5 x^5+\epsilon^6 x^6+\epsilon^8 x^8+\epsilon^9 x^9+\epsilon^{12}x^{12}\right]\,.
\end{equation}
For ${\kgg}=2$ the Coulomb branch is no longer freely generated. The Coulomb branch index reads
\begin{equation}
\mathcal{I}_{\mathbb{Z}_{2},\text{CB}}^{E_6}(x)=\PE\left[x^2+x^{18}+\sum_{j=3}^7x^{2j}-x^{28}\right]\,.
\end{equation}
The generators and relation are 
\begin{equation}
u_2 ,\,u_6,\,u_8,\,\widetilde{u}_1=u_5^2,\,u_{12},\,\widetilde{u}_2=u_5u_9,\,\widetilde{u}_3=u_9^2\,;\quad \widetilde{u}_2^2=\widetilde{u}_1\widetilde{u}_3\,,
\end{equation}
where the $u_j$ are $E_6$-invariant polynomials of degree $j$. The topology is $CB_{E_6,2}\iso\mathbb{C}^4\times\mathbb{C}^2/\mathbb{Z}_2$.
For $\kgg=3,4,6$ the variety is not a complete intersection. To save on lengthy formulas we will list, as an example, only the case of $\kgg=6$. In that case the Coulomb branch index reads
\begin{equation}
\mathcal{I}_{\mathbb{Z}_{6},\text{CB}}^{E_6}(x)=\frac{\left(1 - x^6 + 3 x^{12} + 3 x^{24} +  3 x^{36} - x^{42} + x^{48}\right)\PE\left[6x^6\right]}{(1 + 3 x^6 + 6 x^{12} + 9 x^{18} + 
   11 x^{24} + 11 x^{30} + 9 x^{36} + 6 x^{42} + 3 x^{48} + x^{54})}\,.
\end{equation}

\paragraph{$\mathfrak{g}=E_7$}
By applying \eqref{eqn:SCICBdeg} we have
\begin{equation}
\mathcal{I}_{\mathbb{Z}_{\kgg},\text{CB}}^{E_7}(x)=\frac{1}{|\mathbb{Z}_{\kgg}|}\sum_{\epsilon\in\mathbb{Z}_{\kgg}}\PE\left[\epsilon^2 x^2+\epsilon^6 x^6+\epsilon^{10} x^{10}+\epsilon^{12}x^{12}+\epsilon^{14} x^{14}+\epsilon^{18} x^{18}\right]\,.
\end{equation}
Clearly $\mathcal{I}_{\mathbb{Z}_{1},\text{CB}}^{E_7}(x)=\mathcal{I}_{\mathbb{Z}_{2},\text{CB}}^{E_7}(x)$ and the topology is obviously $CB_{E_7}\iso\mathbb{C}^7$. This is to be expected since $\Out(E_7)$ is trivial. For ${\kgg}=\{3,4,6\}$ we do not get a complete intersection.
% and therefore we can safely conclude that $\mathbb{Z}_n$ discrete gauging of the $\mathfrak{g}=E_7$ theory does not coincide with the $S_{3,1,p'}^7$ S-folds.

\paragraph{$\mathfrak{g}=E_8$}
The Coulomb branch index reads
\begin{equation}
\mathcal{I}_{\mathbb{Z}_{\kgg},\text{CB}}^{E_8}(x)=\frac{1}{|\mathbb{Z}_{\kgg}|}\sum_{\epsilon\in\mathbb{Z}_{\kgg}}\PE\left[\epsilon^2 x^2+\epsilon^8 x^8+\epsilon^{12}x^{12}+\epsilon^{14} x^{14}+\epsilon^{18} x^{18}+\epsilon^{20} x^{20}+\epsilon^{24}x^{24}+\epsilon^{30} x^{30}\right]\,.
\end{equation}
We observe that $\mathcal{I}_{\mathbb{Z}_{1},\text{CB}}^{E_8}(x) = \mathcal{I}_{\mathbb{Z}_{2},\text{CB}}^{E_8}(x)$ and the corresponding topology is $CB_{E_8}\iso\mathbb{C}^8$. Again, this is to be expected due to the fact that $\Out(E_8)=1$. While it's easy to check that for ${\kgg}=3,4,6$ the space is no longer freely generated.

\section{\boldmath Large $N$ limit}\label{sec:largeN}
The large $N$ limit of the index of $G=U(N)$ SYM may be written as \cite{Kinney:2005ej}
\begin{equation}
\mathcal{I}^{\mathfrak{u}(\infty)}(t,y,p,q)=\PE\left[Z^{\text{S.T.}}(t,y,p,q)\right]
\end{equation}
where 
\begin{equation}
Z^{\text{S.T.}}(t,y,p,q)=\sum_{R_2=1}^{\infty}\mathcal{I}_{\mathcal{B}_{[0,R_2,0]}^{\frac{1}{2},\frac{1}{2}}}(t,y,p,q)=\sum_{R_2=1}^{\infty}\sum_{i=0}^{R_2}\mathcal{I}_{\hat{\mathcal{B}}_{[R_2-i,i]}}(t,y,p,q)\,,
\end{equation}
where, in the second line we made us of \eqref{B0m0}.
By applying $r_{\kgg}\newdot s_{\kgg}$ given in \eqref{eqn:Rsym} and \eqref{eqn:Sduality} we can write the single letter index corresponding to the refined index \eqref{eqn:SCIrefined}, it is given by
\begin{equation}
Z^{\text{S.T.}}(t,y,p,q,\epsilon)=\sum_{R_2=1}^{\infty}\sum_{i=0}^{R_2}\epsilon^{R_2-2i}\,\mathcal{I}_{\hat{\mathcal{B}}_{[R_2-i,i]}}(t,y,p,q)\,,
\end{equation}
where we used that
\begin{equation}
(r_k\newdot s_k)\hat{\mathcal{B}}_{[R_1,R_2]}\color{black}=(R_2-R_1)\hat{\mathcal{B}}_{[R_1,R_2]}\color{black}\,.
\end{equation} 
The refined index, at large $N$ is then given by
\begin{equation}
\mathcal{I}^{\mathfrak{u}(\infty)}(t,y,p,q,\epsilon)=\PE\left[Z^{\text{S.T.}}(t,y,p,q,\epsilon)\right]\,.
\end{equation}
The KK supergraviton index graded by $\epsilon$ for $r_k\newdot s_k$, as computed from AdS/CFT, reads \cite{Imamura:2016abe}
\begin{equation}\label{eqn:LargeN}
\begin{aligned}
I^{\text{KK}}\left(t,p,q,y,\epsilon\right)=&\frac{\left(1-\epsilon^{-1}t^3y\right)\left(1-\epsilon^{-1}t^3/y\right)\left(1-t^4\left(\frac{\epsilon}{pq}+\frac{\epsilon p}{q}+\frac{q^2}{\epsilon}\right)+\left(1+\epsilon\right)t^6\right)}{\left(1-t^3y\right)\left(1-t^3/y\right)\left(1-t^2pq/\epsilon\right)\left(1-t^2q/p\epsilon\right)\left(1-t^2\epsilon/q^2\right)}\\
&-\frac{1-\epsilon^{-1}t^6}{\left(1-t^3y\right)\left(1-t^3/y\right)}\,.
\end{aligned}
\end{equation}
Expansion around $t=0$ (we checked up to order $t^{20}$) verifies that
\begin{equation}
Z^{\text{S.T.}}(t,y,p,q,\epsilon)=I^{\text{KK}}\left(t,p,q,y,\epsilon\right)\,.
\end{equation}
The index for the $\mathbb{Z}_{\kgg}$ discrete gauging of the $\mathfrak{u}(N=\infty)$ theory is therefore
\begin{equation}\label{eqn:largeNdis}
\mathcal{I}^{\mathfrak{u}(\infty)}_{\mathbb{Z}_{\kgg}}(t,y,p,q)=\frac{1}{|\mathbb{Z}_{\kgg}|}\sum_{\epsilon\in \mathbb{Z}_{\kgg}}\PE\left[Z^{\text{S.T.}}(t,y,p,q,\epsilon)\right]\,.
\end{equation}

On the other hand, as computed in \cite{Imamura:2016abe}, we may also obtain the index for the $k=1,2,3,4,6$ $\mathcal{N}=3$ S-fold SCFTs at large $N$, $S^{\infty}_{k,\ell}$, by implementing the projection at the level of the single letter index. The spectrum of protected single trace operators in the S-fold $S^{\infty}_{k,\ell}$ theory is given by
\begin{equation}\label{eqn:ZSTk}
Z^{\text{S.T}}_k(t,y,p,q):=\frac{1}{|\mathbb{Z}_k|}\sum_{R_2=1}^{\infty}\sum_{i=0}^{R_2}\sum_{\epsilon\in\mathbb{Z}_k}\epsilon^{R_2-2i}\mathcal{I}_{\hat{\mathcal{B}}_{[R_2-i,i]}}(t,y,p,q)\,,
\end{equation}
for $k=1,2,3,4,6$.
Note that, at large $N$, the index does not distinguish between theories with different values of $\ell$ \cite{Imamura:2016abe}.
One advantage of \eqref{eqn:ZSTk} is that it manifestly organises expression into multiplets of $\mathfrak{su}(2,2|3)$. The index for the S-fold at large $N$ is then given by
\begin{equation}
\mathcal{I}^{N=\infty}_{\mathbb{Z}_k\text{ S-fold}}(t,p,q,y)=\PE\left[Z_k^{\text{S.T}}(t,p,q,y)\right]\,.
\end{equation}
Note that the procedure \eqref{eqn:ZSTk} is a S-fold and not a discrete gauging since it is implemented at the level of the single particle index.
It is clear that 
\begin{equation}
\mathcal{I}^{\mathfrak{u}(\infty)}_{\mathbb{Z}_{{\kgg}=k}}(t,y,p,q)\neq\mathcal{I}^{N=\infty}_{\mathbb{Z}_k\text{ S-fold}}(t,p,q,y)\,.
\end{equation}

\section{Conclusions}

In this paper we gave a prescription on how to implement the discrete gauging of a four dimensional $\mathcal{N}=4$ mother theory, resulting in a $\mathcal{N}=3$ daughter theory, at the level of the superconformal index.
We explicitly computed  the Coulomb branch limit of the index as well as the Higgs branch Hilbert series for a number of  theories based on simply laced groups.
For rank one theories, the Coulomb branch index and Higgs branch Hilbert series we computed reproduce precisely all known results, while, for higher rank theories we make concrete predictions for the  Coulomb and Higgs branches of these $\mathcal{N}=3$ theories.
Most strikingly we find that, in general, the higher rank theories, have non-freely generated Coulomb branches. In a few cases the Coulomb branch is a complete intersection variety and, using the Coulomb branch index, we were able to read off the topology of the corresponding space. Generally the Coulomb branch of the theory after the discrete gauging is not a complete intersection and the topology becomes harder to extract. It would be interesting to further study this aspect in the future. 
%To our knowledge the mathematics that are needed to solve this problem are now quite developed.

Since the superconformal index of the $\mathfrak{u}(1)$ theory is easily reorganised into $\mathcal{N}=3$ multiplets we were able to compute the full superconformal index for the discrete gauging of it, given in equation \eqref{eqn:u1index}. Moreover, in the large $N$ limit, a similar reorganisation happens meaning that it is also possible to compute the superconformal index for the discrete gauging, given in equation \eqref{eqn:largeNdis}. For general rank the computation of the full index, or other more refined limits such as the Schur limit, is much more difficult and we leave it for future work.
However, we can easily compute the Coulomb branch limit of the superconformal index and
the Higgs branch Hilbert series. Other, more refined, limits contain more types of short multiplets,
which of course contain more interesting information. In particular the Schur index is related to the vacuum character of chiral algebras.
The latter 
 allows for the computation of correlation functions  in a protected  sector 
\cite{Beem:2013sza}. 
For $\mathcal{N}=3$ theories the study of chiral algebras  was  initiated  in \cite{Nishinaka:2016hbw,Lemos:2016xke} and it would be very interesting to further pursue.
With the help of the superconformal index, we can construct and analyse the corresponding chiral algebras for the discrete gauging that we studied in this paper. %This work is being carried out in \cite{Ourselves}.

%In \cite{Gadde:2009kb,Gadde:2011ik} it was demonstrated that the superconformal index of $\mathcal{N}=4$ SYM may be interpreted as a correlation function in a TQFT living on $T^2$. It seems reasonable to assume that the $\mathcal{N}=3$ theory obtained via discrete gauging also inherits this structure. It would be interesting to further investigate this point in the future.

It is important to note that the spectrum of non-local operators may reduce the possible $\mathbb{Z}_{\kgg}$'s that can enhance to symmetries of the theory and therefore be gauged. The standard superconformal index that we studied in this paper can say nothing about the non-local operator spectrum.  It captures only the spectrum of protected local operators.
Using our current tools we only claim that if a theory exists  we can compute its index, but we have no way of deciding if a theory actually exists.
To break this impasse, a very interesting quantity to compute for the theories obtained via discrete gauging is the Lens space index \cite{Benini:2011nc,Alday:2013rs,Razamat:2013jxa,Razamat:2013opa}. It is a generalisation of the standard superconformal index that has a representation as a path integral on $\mathbb{S}^1\times \mathbb{S}^3/\mathbb{Z}_r$. For $r=1$ this reduces to the usual superconformal index, however, since $\pi_1\left(\mathbb{S}^3/\mathbb{Z}_r\right)=\mathbb{Z}_r$ the Lens space index has the advantage that it is sensitive to the spectrum of line operators of the theory.
Our construction can be immediately generalized for $r\neq1$.
 In a similar spirit it is also possible to compute the index in the presence of certain extended operators \cite{Gaiotto:2012xa,Gang:2012yr}.
 These should also shed light to the possible discrete gaugings allowed for a given theory.
Computing such quantities may be able to teach us more about, the currently mysterious, `new' $\mathcal{N}=4$ theories \cite{Argyres:2016yzz} and discrete gaugings thereof.

Finally, our procedure can also be applied to discrete gauging that preserves $\mathcal{N}=2$ superconformal symmetry as in \cite{Argyres:2016yzz} and will most definitely help us discover their novel properties.

\acknowledgments

We are very grateful to Philip Argyres and Mario Martone for having shared with us a draft of their article and for useful discussions. We are also grateful to Madalena Lemos for many important discussions and for reading and commenting on the draft of our article.
It is also a great pleasure to thank Antoine Bourget and Troy Figiel for interesting discussions.
Our work is supported by the German Research Foundation (DFG) via the Emmy
Noether program ``Exact results in Gauge theories''.

\bigskip

\appendix

\section{The preserved superconformal algebra}\label{sec:preservedSCA}
\paragraph{Even subalgebra}
The even subalgebra of $\mathfrak{psu}(2,2|4)$ is $\mathfrak{b}=\mathfrak{so}(4,2)\oplus\mathfrak{su}(4)$ which we take to be generated by $M^{\mu\nu},K_{\mu},P^{\mu},E$ with $\mu,\nu=1,2,3,4$ and $R_I^J$, $I,J=1,2,3,4$. The Cartans of $\mathfrak{su}(4)$ are $R_i=R_i^i-R^{i+1}_{i+1}$ with $i=1,2,3$. We wish to discuss which generators are preserved by the S-folding/discrete gauging procedure.
Recall that $SL(2,\mathbb{Z})$ transformations can be defined such that they commute with the generators of $\mathfrak{b}$ \cite{Kapustin:2006pk}. In particular $[s_k,\mathfrak{b}]=0$. Hence $s_k$ acts non-trivially only on the fermionic subalgebra which we will discuss momentarily. Hence the subalegbra of $\mathfrak{b}$ preserved by the S-folding/discrete gauging is simply the centraliser of $r_k=\frac{R_1}{2}+R_2+\frac{3R_3}{2}=\frac{1}{2}\sum_{i=1}^3R_i^i-\frac{3}{2}R_4^4$ modulo $k$ in $\mathfrak{b}$. Clearly $\left[r_k,\mathfrak{so}(4,2)\right]=0$. On the other hand, using $\left[R_I^J,R^P_Q\right]=\delta_Q^JR_I^P-\delta^P_IR_Q^J$ it can be shown that
\begin{equation}
[r_k,R^J_I]=\begin{cases}
0&I,J\in\{1,2,3\}\,,\\
0&I=J=4\,,\\
2R^4_I&I\in\{1,2,3\}\,,J=4\,,\\
-2R_4^J&I=4\,,J\in\{1,2,3\}\,.
\end{cases}
\end{equation}
Therefore, the subalgebra of $\mathfrak{su}(4)$ preserved by $r_{k\geq3}$ are given by the $R_I^J$ with $I,J=1,2,3$ and $R_4^4$. These generators span a $\mathfrak{su}(3)\oplus\mathfrak{u}(1)$ algebra. Note however that, since we quotient by $e^{\frac{2\pi\iu}{k}r_k\newdot s_k}$, when $k=1,2$ the full $\mathfrak{su}(4)$ is preserved.

\paragraph{Odd subalgebra} 
The odd subalgebra of $\mathfrak{psu}(2,2|4)$ is spanned by nilpotent generators (supercharges) which sit in representations of the bosonic subalgebra $\mathfrak{b}$. Any representation of $\mathfrak{b}$ can be decomposed into representations of a maximal compact subalgebra $\mathfrak{u}(1)_E\oplus\mathfrak{su}(2)_1\oplus \mathfrak{su}(2)_2\oplus \mathfrak{su}(4)$. The supercharges are then given by
\begin{equation}
\mathcal{Q}_{\alpha}^I\in\left(\frac{1}{2},\mathbf{2},\mathbf{1},\mathbf{4}\right)\,,\quad \widetilde{\mathcal{Q}}_{\dot\alpha I} \in\left(\frac{1}{2},\mathbf{1},\mathbf{2},\mathbf{\overline{4}}\right)\,,\quad \mathcal{S}^{\alpha}_I\in\left(-\frac{1}{2},\mathbf{\overline{2}},\mathbf{1},\mathbf{\overline{4}}\right)\,,\quad \widetilde{\mathcal{S}}^{\dot\alpha I}\in\left(-\frac{1}{2},\mathbf{1},\mathbf{\overline{2}},\mathbf{4}\right)\,.
\end{equation}
The action on the supercharges is then given by
\begin{gather}
[r_k,\mathcal{Q}^{I}_{\alpha}] = \begin{cases}
\mathcal{Q}^{I}_{\alpha} & I=1,2,3\\
-3\mathcal{Q}^{4}_{\alpha} &I=4
\end{cases}\,,\quad[r_k,\widetilde{\mathcal{Q}}_{\dot\alpha I}]=\begin{cases}
-\widetilde{\mathcal{Q}}_{\dot\alpha I} & I=1,2,3\\
3\widetilde{\mathcal{Q}}_{\dot\alpha 4} &I=4
\end{cases} \, ,\\
[r_k,\mathcal{S}_{I}^{\alpha}]=\begin{cases}
-\mathcal{S}^{I}_{\alpha} & I=1,2,3\\
3\mathcal{S}^{4}_{\alpha} &I=4
\end{cases}\,,\quad[r_k,\widetilde{\mathcal{S}}^{\dot\alpha I}] = \begin{cases}
\widetilde{\mathcal{S}}_{\dot\alpha I} & I=1,2,3\\
-3\widetilde{\mathcal{S}}_{\dot\alpha 4} &I=4
\end{cases} \, ,
\end{gather}
On the other hand, $s_k$ acts on the supercharges by \cite{Kapustin:2006pk,Garcia-Etxebarria:2015wns,Garcia-Etxebarria:2016erx}
\begin{equation}
[s_k,\mathcal{Q}^{I}_{\alpha}]=-\mathcal{Q}^{I}_{\alpha} \,,\quad  [s_k,\widetilde{\mathcal{Q}}_{\dot\alpha I}]=\widetilde{\mathcal{Q}}_{\dot\alpha I} \, ,\quad [s_k,\mathcal{S}_{I}^{\alpha}]=\mathcal{S}_{I}^{\alpha} \,,\quad  [r_k,\widetilde{\mathcal{S}}^{\dot\alpha I} ]=-\widetilde{\mathcal{S}}_{\dot\alpha I} \,.
\end{equation}
Therefore, for $k\geq 3$, quotienting by $e^{\frac{2\pi\iu}{k}(r_k\newdot s_k)}\in\mathbb{Z}_k$ preserves 12 Poincar\'e supercharges and 12 conformal supercharges giving rise to $\mathcal{N}=3$ superconformal symmetry in four dimensions.
All in all, for $k\geq3$, a full $\mathfrak{su}(2,2|3)\subset\mathfrak{psu}(2,2|4)$ superconformal algebra is preserved.

\section{\boldmath Indices for $\mathfrak{su}(2,2|2)$ multiplets}\label{sec:indicies}
\renewcommand{\arraystretch}{1.2}
Long multiplets $\mathcal{A}^{E}_{R,r,(j_1,j_2)}$ are generic, unitary, modules of the $\mathfrak{su}(2,2|2)$ superconformal algebra. The multiplets are labelled by the values of the highest weight state (superconformal primary) $(E,R,r,j_1,j_2)$ under the maximal bosonic subalgebra \eqref{eqn:su222maximal}. When the some of representation labels take on certain values the superconformal primary is annihilated by (linear combinations of) some of the supercharges $\mathcal{Q}_{\alpha}^I$, $\widetilde{\mathcal{Q}}_{\dot{\alpha} I}$ and the multiplet is said to be shortened. The superconformal index \eqref{eqn:SCI} counts short multiplets modulo those that can recombine into long multiplets. The recombination rules are given by \cite{DolanOsborn}
\begin{align}
&\mathcal{A}_{R,r,(j_1,j_2)}^{2R+r+2j_1+2}\iso \mathcal{C}_{R,r,(j_1,j_2)}\oplus\mathcal{C}_{R+\frac{1}{2},r+\frac{1}{2},\left(j_1-\frac{1}{2},j_2\right)}\label{eqn:short1} \, , \\
&\mathcal{A}_{R,r,(j_1,j_2)}^{2R-r+2j_2+2}\iso \overline{\mathcal{C}}_{R,r,(j_1,j_2)}\oplus\overline{\mathcal{C}}_{R+\frac{1}{2},r-\frac{1}{2},\left(j_1,j_2-\frac{1}{2}\right)} \, , \\
&\mathcal{A}_{R,j_1-j_2,(j_1,j_2)}^{2R+j_1+j_2+2}\iso \hat{\mathcal{C}}_{R,(j_1,j_2)}\oplus\hat{\mathcal{C}}_{R+\frac{1}{2},\left(j_1-\frac{1}{2},j_2\right)}\oplus\hat{\mathcal{C}}_{R+\frac{1}{2},\left(j_1,j_2-\frac{1}{2}\right)}\oplus\hat{\mathcal{C}}_{R+1,\left(j_1-\frac{1}{2},j_2-\frac{1}{2}\right)}\,.
\end{align}
By allowing the $j_1,j_2$ to take on the value $-1/2$ we can write
\begin{align}
\mathcal{C}_{R,r,\left(-\frac{1}{2},j_2\right)}\iso\mathcal{B}_{R+\frac{1}{2},r+\frac{1}{2},(0,j_2)}\,,\quad&\overline{\mathcal{C}}_{R,r,\left(j_1,-\frac{1}{2}\right)}\iso\overline{\mathcal{B}}_{R+\frac{1}{2},r-\frac{1}{2},(j_1,0)}\,,\label{eqn:Semishort}\\
\hat{\mathcal{C}}_{R,\left(-\frac{1}{2},j_2\right)}\iso\mathcal{D}_{R+\frac{1}{2},(0,j_2)}\,,\quad & \hat{\mathcal{C}}_{R,\left(j_1,-\frac{1}{2}\right)}\iso\overline{\mathcal{D}}_{R+\frac{1}{2},(j_1,0)}\,,\\
\hat{\mathcal{C}}_{R,\left(-\frac{1}{2},-\frac{1}{2}\right)}\iso\mathcal{D}_{R+\frac{1}{2},\left(0,-\frac{1}{2}\right)}\iso&\, \overline{\mathcal{D}}_{R+\frac{1}{2},\left(-\frac{1}{2},0\right)}\iso\hat{\mathcal{B}}_{R+1}\label{eqn:short2}\,,
\end{align}
for $R\geq0$. Equations \eqref{eqn:short1}-\eqref{eqn:short2} constitute the most general recombination rules for any unitary $\mathcal{N}=2$ SCFT.
We summarize in Table \ref{tab:short} the different shortening conditions.

\begin{table}
\begin{center}
\begin{tabular}{|c|c|c|c|c|}
\hline
\multicolumn{4}{|c|}{Shortening Conditions} & Multiplet \\
\hline
\hline
$\mathcal{B}_1$ & $\mathcal{Q}_{1\alpha}|R,r \rangle^{h.w.}=0$ & $j_1=0$ & $E=2R+r$ & $\mathcal{B}_{R,r(0,j_2)}$\\
\hline
$\bar{\mathcal{B}}_2$ & $\tilde{\mathcal{Q}}_{2\dot{\alpha}}|R,r \rangle^{h.w}=0$ & $j_2=0$ & $E=2R-r$ & $\bar{\mathcal{B}}_{R,r(j_1,0)}$\\
\hline
$\mathcal{E}$ & $\mathcal{B}_1 \cap \mathcal{B}_2 $ & $R=0$ & $E=r$ & $\mathcal{E}_{r(0,j_2)}$\\
\hline
$\mathcal{\bar{E}}$ & $\bar{\mathcal{B}}_1 \cap \bar{\mathcal{B}}_2$ & $R=0$ & $E=-r$ & $\mathcal{\bar{E}}_{r(j_1,0)}$\\
\hline
$\hat{\mathcal{B}}$ & $\mathcal{B}_1 \cap \bar{\mathcal{B}}_2$ & $r=0, j_1,j_2=0$ & $E=2R$ & $\hat{\mathcal{B}}_{R}$\\
\hline
\hline
$\mathcal{C}_1$ & $\epsilon^{\alpha\beta}\mathcal{Q}_{1\beta}|R,r\rangle_{\alpha}^{h.w.} = 0 $ &  & $E=2+2j_1+2R+r$  & $\mathcal{C}_{R,r(j_1,j_2)}$\\
& $(\mathcal{Q}_{1})^2|R,r \rangle^{h.w.}=0 \ \textrm{for} \ j_1=0$ & & $E=2+2R+r$ & $\mathcal{C}_{R,r(0,j_2)}$\\
\hline
$\mathcal{\bar{C}}_2$ & $\epsilon^{\dot{\alpha}\dot{\beta}}\tilde{\mathcal{Q}}_{2\dot{\beta}}|R,r\rangle_{\dot{\alpha}}^{h.w.} = 0 $ &  & $E=2+2j_2+2R-r$  & $\mathcal{\bar{C}}_{R,r(j_1,j_2)}$\\
& $(\mathcal{\tilde{Q}}_{2})^2|R,r \rangle^{h.w.}=0 \ \textrm{for} \ j_2=0$ & & $E=2+2R-r$ & $\mathcal{\bar{C}}_{R,r(j_1,0)}$\\
\hline
& $\mathcal{C}_1 \cap \mathcal{C}_2$ & $R=0$ & $E=2+2j_1+r$ & $\mathcal{C}_{0,r(j_1,j_2)}$\\
\hline
& $\mathcal{\bar{C}}_1 \cap \mathcal{\bar{C}}_2$ & $R=0$ & $E=2+2j_2-r$ & $\mathcal{\bar{C}}_{0,r(j_1,j_2)}$\\
\hline
$\mathcal{\hat{C}}$ & $\mathcal{C}_1 \cap \bar{\mathcal{C}}_2 $ & $r=j_2 -j_1$ & $E=2+2R+j_1+j_2$  & $\mathcal{\hat{C}}_{R(j_1,j_2)}$\\
\hline
& $\mathcal{C}_1 \cap \mathcal{C}_2 \cap \mathcal{\bar{C}}_1 \cap \mathcal{\bar{C}}_2$ & $R=0,r=j_2-j_1$ & $E=2+j_1+j_2$ & $\mathcal{\hat{C}}_{0(j_1,j_2)}$\\
\hline
\hline
$\mathcal{D}$ & $\mathcal{B}_1 \cap \mathcal{\bar{C}}_2$ & $r=j_2+1$ & $E=1+2R+j_2$ & $\mathcal{D}_{R(0,j_2)}$\\
\hline
$\mathcal{\bar{D}}$ & $\mathcal{\bar{B}}_2 \cap \mathcal{C}_1$ & $-r=j_1+1$ & $E=1+2R+j_1$ & $\mathcal{\bar{D}}_{R(j_1,0)}$\\
\hline
& $\mathcal{E} \cap \mathcal{\bar{C}}_2$ & $r=j_2+1, R=0$ & $E=r=1+j_2$ & $D_{0,(0,j_2)}$\\
\hline
& $\mathcal{\bar{E}} \cap \mathcal{C}_1$ & $-r = j_1+1,R=0$ & $E=-r=1+j_1$ & $\mathcal{\bar{D}}_{0,(j_1,0)}$ \\
\hline
\end{tabular}
\caption{Shortening conditions and short multiplets for the $\mathcal{N}=2$ SCA.\label{tab:short}}
\end{center}
\end{table}

We have that
\begin{align}
\label{eq:MultiIndexE}
&\mathcal{I}_{\mathcal{E}_{r,(0,j_2)}}=(-1)^{2j_2}t^{2r}(pq)^r\frac{1-t(pq)^{-1}\chi_1(y)+t^2(pq)^{-2}}{(1-t^3y)(1-t^3y^{-1})}\chi_{2j_2}(y)\quad r\geq2\,, \\
&\mathcal{I}_{\mathcal{D}_{0,(0,j_2)}}=(-1)^{2j_2}\frac{pqt^2\chi_{2j_2}(y)-t^3\chi_{2j_2+1}(y)-t^5pq\chi_{2j_2-1}(y)+t^6\chi_{2j_2}(y)}{(1-t^3y)(1-t^3y^{-1})}\,,\\
&\mathcal{I}_{\overline{\mathcal{D}}_{0,(j_1,0)}}=(-1)^{2j_1+1}\frac{t^{4j_1+4}}{(pq)^{j_1+1}}\frac{1-(pq)t^2}{(1-t^3y)(1-t^3y^{-1})}\,,\\
&\mathcal{I}_{\mathcal{C}_{R,r(j_1,j_2)}}=(-1)^{2j_1+2j_2+1}\frac{t^{4+4R+6j_1+2r}}{(pq)^{R+1-r}}\frac{\left(1-t^2pq\right)\left(t^2pq-t^3\chi_1(y)+\frac{t^4}{pq}\right)}{\left(1-t^3y\right)\left(1-t^3y^{-1}\right)}\chi_{2j_2}(y)\,,\\
&\mathcal{I}_{\hat{\mathcal{C}}_{R(j_1,j_2)}}=(-1)^{2j_1+2j_2}\frac{t^{6+4R+4j_1+2j_2}}{(pq)^{R+j_1-j_2}}\frac{\left(1-t^2pq\right)\left(\frac{t}{pq}\chi_{2j_2+1}(y)-\chi_{2j_2}(y)\right)}{(1-t^3y)(1-t^3y^{-1})}\,,\\
%\label{eq:MultiIndex0}
&\mathcal{I}_{\overline{\mathcal{E}}_{r,(j_1,0)}}=\mathcal{I}_{\mathcal{E}_{0,(0,0)}}=\mathcal{I}_{\overline{\mathcal{C}}_{R,r(j_1,j_2)}}=\mathcal{I}_{\mathcal{A}^{E}_{R,r(j_1,j_2)}}=0\label{eqn:index0}\,.
\end{align}
These may be obtained from \cite{Gadde:2011uv} by conjugation (exchanging $r\to-r$, $j_1\leftrightarrow j_2$) and setting $\tau=t^2(pq)^{-1/2}$, $\sigma=ty(pq)^{1/2}$, $\rho=ty^{-1}(pq)^{1/2}$).
By applying \eqref{eqn:Bnmdecomp}-\eqref{eqn:B0n} in combination with \eqref{eqn:Semishort}-\eqref{eqn:index0} one can compute the contribution to the index of the $\mathfrak{su}(2,2|3)$ multiplets $\hat{\mathcal{B}}_{[R_1,R_2]}$.

%\section{Hall-Littlewood limit of the index}
%\subfile{appendixHL}

\bibliography{biblio}
\bibliographystyle{JHEP}

\end{document}